\documentclass[prd,preprintnumbers,twocolumn,amsmath,nofootinbib,amssymb]{revtex4}

\usepackage{graphicx,color,dcolumn,booktabs,bm}
\usepackage{longtable,lscape}
\usepackage{makecell}
\usepackage{txfonts}
\usepackage{overpic}
\usepackage{amssymb}
\usepackage{epstopdf}
\usepackage{indentfirst}
\usepackage{feynmf}   
\usepackage{slashed}  
\usepackage{cases}
\usepackage{color}
\usepackage{float}
\usepackage{multirow}
\usepackage{ulem}
\usepackage{enumerate}
\usepackage{graphicx,color,dcolumn,booktabs,bm}
\usepackage{epsfig,dsfont,amssymb,amsmath,amsfonts,amsbsy,mathrsfs}

\graphicspath{{Figures/}} %

\usepackage{hyperref}
\hypersetup{colorlinks,citecolor=blue,anchorcolor=red,menucolor=red, linkcolor=red,filecolor=red,runcolor=red,urlcolor=blue,frenchlinks=true}

\makeatletter
\@addtoreset{equation}{section}
\makeatother

\allowdisplaybreaks

\begin{document}

\title{Singly-strange hidden-bottom and double-bottom molecular pentaquarks}
\author{Fu-Lai Wang$^{1,2,3,4}$}
\email{wangfulai@lzu.edu.cn}
\author{Jia-Ning Cui$^{1}$}
\author{Rui Chen$^{5,6}$}
\email{chenrui@hunnu.edu.cn}
\author{Xiang Liu$^{1,2,3,4}$\footnote{Corresponding author}}
\email{xiangliu@lzu.edu.cn}
\affiliation{$^1$School of Physical Science and Technology, Lanzhou University, Lanzhou 730000, China\\
$^2$Lanzhou Center for Theoretical Physics, Key Laboratory of Theoretical Physics of Gansu Province, Key Laboratory of Quantum Theory and Applications of MoE, Gansu Provincial Research Center for Basic Disciplines of Quantum Physics, Lanzhou University,
Lanzhou 730000, China\\
$^3$MoE Frontiers Science Center for Rare Isotopes, Lanzhou University, Lanzhou 730000, China\\
$^4$Research Center for Hadron and CSR Physics, Lanzhou University and Institute of Modern Physics of CAS, Lanzhou 730000, China\\
$^5$Key Laboratory of Low-Dimensional Quantum Structures and Quantum Control of Ministry of Education, Department of Physics and Synergetic Innovation Center for Quantum Effects and Applications, Hunan Normal University, Changsha 410081, China\\
$^6$Hunan Research Center of the Basic Discipline for Quantum Effects and Quantum Technologies, Hunan Normal University, Changsha 410081, China}

\begin{abstract}
Inspired by the observation of the singly-strange hidden-charm pentaquarks $P_{\psi s}^{\Lambda}(4459)$ and $P_{\psi s}^{\Lambda}(4338)$, we present a systematic investigation of the mass spectra of the singly-strange hidden-bottom and double-bottom molecular pentaquarks arising from the $\Xi_b^{(\prime,\,*)}B^{(*)}$ and $\Xi_b^{(\prime,\,*)}\bar B^{(*)}$ interactions. Using the one-boson-exchange model, we derive the effective potentials for the $\Xi_b^{(\prime,\,*)}B^{(*)}$ systems by exchanging light mesons. The effective potentials for the corresponding $\Xi_b^{(\prime,\,*)}\bar B^{(*)}$ systems are then obtained via the $G$-parity rule. By solving the coupled-channel Schr\"odinger equation with $S$-$D$ wave mixing and coupled-channel effects, we search for loosely bound states with all possible isospin and spin-parity quantum numbers. Our results show that the $\Xi_b B$ state with $I(J^P)=0(1/2^-)$, the $\Xi_b B^*$ states with $I(J^P)=0(1/2^-)$ and $0(3/2^-)$, the $\Xi_b^\prime B$ state with $I(J^P)=0(1/2^-)$, the $\Xi_b^* B$ state with $I(J^P)=0(3/2^-)$, the $\Xi_b^\prime B^*$ states with $I(J^P)=0(1/2^-)$ and $0(3/2^-)$, and the $\Xi_b^* B^*$ states with $I(J^P)=0(1/2^-)$, $0(3/2^-)$, and $0(5/2^-)$ are identified as promising candidates of the singly-strange hidden-bottom molecular pentaquarks. Their $\Xi_b^{(\prime,\,*)}\bar B^{(*)}$ partners are likewise predicted to be the singly-strange double-bottom molecular pentaquark candidates. These predictions provide concrete targets for future experimental searches for exotic hadrons in the bottom sector, for example at LHCb.
\end{abstract}
\maketitle

\section{Introduction}\label{introduction}

The study of exotic hadronic states beyond the conventional hadron remains one of the most active research directions in hadron physics. The discovery of the $X(3872)$ by Belle in 2003 stimulated extensive experimental and theoretical activity \cite{Belle:2003nnu}, yielding a series of exotic hadronic candidates including the charmonium-like states $XYZ$, the hidden-charm pentaquarks $P_{\psi}^{N}$, their strange counterparts $P_{\psi s}^{\Lambda}$, and the double-charm tetraquark $T_{cc}^+$ \cite{Liu:2013waa,Hosaka:2016pey,Chen:2016qju,Richard:2016eis,Lebed:2016hpi,Liu:2019zoy,Brambilla:2019esw,Chen:2022asf,Olsen:2017bmm,Guo:2017jvc,Meng:2022ozq,Liu:2024uxn,Wang:2025sic,Wang:2025dur,Dai:2026fkg,Bai:2026atm}. These observations have reshaped our understanding of strong interactions in the nonperturbative regime and have motivated extensive investigations into the inner structures of exotic hadronic states. Among the various theoretical approaches, the hadronic molecular picture has been particularly successful. In this framework, a number of newly observed hadronic states are interpreted as the loosely bound states of two conventional hadrons, analogous to the deuteron in nuclear physics. This picture is particularly suitable in sectors involving the heavy quarks, where the nonrelativistic approximation is reliable and the heavy-quark symmetry provides constraints on the interactions between hadrons \cite{Liu:2024uxn}.

\begin{figure*}[htbp]
\includegraphics[scale=0.53]{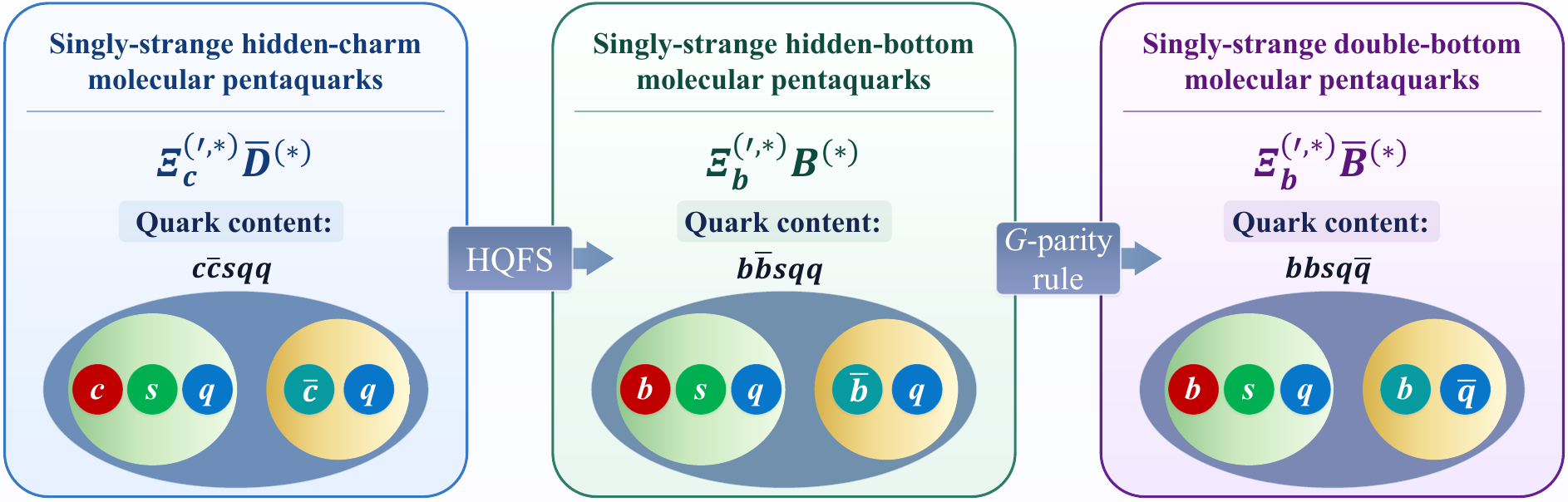}
\caption{Building upon the discovery of the singly-strange hidden-charm pentaquarks \cite{LHCb:2020jpq,LHCb:2022ogu}, we extend the molecular paradigm to the hidden-bottom and double-bottom sectors, and systematically predict the mass spectra of the $\Xi_b^{(\prime,\,*)}B^{(*)}$ and $\Xi_b^{(\prime,\,*)}\bar B^{(*)}$ molecular states.}\label{Expansion}
\end{figure*}

The experimental study of the heavy pentaquark states entered a new phase in 2015, when LHCb announced the observation of two resonance structures, $P_{\psi}^N(4380)$ and $P_{\psi}^N(4450)$, in the $J/\psi p$ invariant-mass spectrum from the $\Lambda_b^0 \to J/\psi p K^-$ decays \cite{Aaij:2015tga}. Subsequent analyses utilizing the high-statistics data from Run I and II further refined this observation. The $P_{\psi}^N(4450)$ peak was resolved into two narrower states, $P_{\psi}^N(4440)$ and $P_{\psi}^N(4457)$, while a lower-mass resonance, $P_{\psi}^N(4312)$, was also identified \cite{Aaij:2019vzc}. The masses of these observed hidden-charm pentaquarks lie close to the $\Sigma_c\bar D^{(*)}$ thresholds, thereby providing evidence for their interpretation as the hadronic molecules \cite{Li:2014gra,Karliner:2015ina,Wu:2010jy,Wang:2011rga,Yang:2011wz,Wu:2012md,Chen:2015loa}. A natural extension of this family is the inclusion of strange quark. In 2020, LHCb reported evidence for a hidden-charm pentaquark with strangeness, $P_{\psi s}^{\Lambda}(4459)$, observed in the $J/\psi\Lambda$ invariant-mass spectrum of the $\Xi_b^- \to J/\psi \Lambda K^-$ decays \cite{LHCb:2020jpq}. This was followed in 2022 by the observation of another singly-strange hidden-charm pentaquark, $P_{\psi s}^{\Lambda}(4338)$, in the $J/\psi\Lambda$ channel of the $B^- \to J/\psi \Lambda \bar p$ decays \cite{LHCb:2022ogu}. These findings have prompted a wealth of theoretical studies. In particular, $P_{\psi s}^{\Lambda}(4459)$ has been widely discussed as a $\Xi_c\bar D^*$ molecular state, while $P_{\psi s}^{\Lambda}(4338)$ is interpreted as a $\Xi_c\bar D$ molecular candidate \cite{Wu:2010vk,Weng:2019ynv,Wang:2019nvm,Xiao:2019gjd,Anisovich:2015zqa,Chen:2016ryt,Hofmann:2005sw}. In 2025, Belle reported evidence for the $P_{\psi s}^{\Lambda}(4459)$ in the $\Upsilon(1S,\,2S)$ inclusive decays \cite{Belle:2025pey}, adding further experimental input and calling for systematic theoretical explorations of the pentaquark states across various quark-flavor configurations.

Although the charm sector of the pentaquark spectroscopy has seen remarkable progress in recent years, the hidden-bottom pentaquark states remain largely uncharted territory both theoretically and experimentally \cite{Dai:2026fkg}. Nevertheless, as illustrated in Fig. \ref{Expansion}, the heavy-quark flavor symmetry (HQFS) suggests a natural correspondence between the hidden-charm and hidden-bottom systems \cite{Liu:2024uxn}. For every observed or predicted singly-strange hidden-charm molecular pentaquark, a bottom analog is expected to exist. In fact, replacing the charm quark with the heavier bottom quark brings several theoretical advantages. The larger bottom quark mass makes the heavy-quark spin symmetry (HQSS) more reliable, since the symmetry-breaking corrections scale as $\Lambda_{\rm QCD}/m_{Q}$ and are therefore significantly smaller than in the charm case \cite{Neubert:1993mb, Wise:1992hn}. Additionally, the heavier bottom quark reduces the kinetic energy of the discussed system, which not only improves the accuracy of the nonrelativistic approximation but also strengthens the inter-hadron binding \cite{Li:2014gra,Karliner:2015ina}. On the experimental side, the prospects for discovering the hidden-bottom pentaquarks are improving \cite{Bediaga:2018lhg,Dai:2026fkg}. With the ongoing accumulation of Run III data at LHCb and the forthcoming High-Luminosity LHC program, the study of these states in final states such as $\Upsilon \Lambda$ is expected to become an appealing and promising direction for future research.

Furthermore, the discovery of the double-charm tetraquark $T_{cc}^+(3875)$ by LHCb in 2021 marked a milestone in the exotic hadron spectroscopy \cite{LHCb:2021vvq,LHCb:2021auc}, providing the first clear evidence for a doubly heavy multiquark state. With its mass lying below the $D^0 D^{*+}$ threshold, $T_{cc}^+(3875)$ strongly favors a molecular interpretation \cite{Manohar:1992nd,Ericson:1993wy,Tornqvist:1993ng,Janc:2004qn,Ding:2009vj,Molina:2010tx,Ding:2020dio,Li:2012ss,Xu:2017tsr,Liu:2019stu,Ohkoda:2012hv,Tang:2019nwv}, and has since inspired extensive theoretical studies on doubly heavy multiquark systems, including the double-charm molecular pentaquarks. This progress not only validates the molecular picture in the doubly heavy sector but also raises the prospect of discovering analogous states with the bottom quarks. Motivated by these developments, we extend the investigation to the singly-strange double-bottom molecular pentaquarks formed from the $\Xi_b^{(\prime,\,*)}$ baryons and the $\bar B^{(*)}$ mesons, aiming to provide a systematic study of their binding properties and mass spectrum.

In this work, we present a systematic investigation of the mass spectra of two novel classes of the molecular pentaquark states. The first class consists of the singly-strange hidden-bottom molecular pentaquarks formed from the $\Xi_b^{(\prime,\,*)}$ baryons and the $B^{(*)}$ mesons, carrying the quark content $b\bar b s q q$ ($q=u~{\rm or}~d$). These states are the bottom counterparts of the observed singly-strange hidden-charm pentaquarks $P_{\psi s}^{\Lambda}(4459)$ and $P_{\psi s}^{\Lambda}(4338)$ \cite{LHCb:2020jpq,LHCb:2022ogu}. The second class involves the singly-strange double-bottom molecular pentaquarks arising from the $\Xi_b^{(\prime,\,*)}\bar B^{(*)}$ interactions, with the quark content $bbsq\bar q$, representing a new category of the doubly heavy molecular pentaquarks. To describe the interactions between the $\Xi_b^{(\prime,\,*)}$ baryons and the $B^{(*)}$ mesons, we adopt the one-boson-exchange (OBE) model, which has been a reliable and predictive framework for studying the heavy-flavor hadronic molecules \cite{Chen:2016qju}. This model has been successfully applied to explain the observed hidden-charm pentaquarks, the double-charm tetraquark, and many other exotic candidates \cite{Liu:2013waa,Hosaka:2016pey,Chen:2016qju,Richard:2016eis,Lebed:2016hpi,Liu:2019zoy,Brambilla:2019esw,Chen:2022asf,Olsen:2017bmm,Guo:2017jvc,Meng:2022ozq,Liu:2024uxn,Wang:2025sic,Wang:2025dur,Dai:2026fkg,Bai:2026atm}. The extension of the OBE model to the $\Xi_b^{(\prime,\,*)}B^{(*)}$ systems is both straightforward and well justified. As shown in Fig. \ref{Expansion}, the interactions for the $\Xi_b^{(\prime,\,*)}\bar B^{(*)}$ systems are then obtained through the $G$-parity rule \cite{Klempt:2002ap}, which provide a consistent connection between the two sectors. By solving the coupled-channel Schr\"odinger equation with the $S$-$D$ wave mixing and coupled-channel effects, we investigate the bound-state properties for all possible isospin and spin-parity quantum numbers, providing a comprehensive survey of the mass spectra of the singly-strange molecular pentaquarks in both the hidden-bottom and double-bottom sectors.

The remainder of this paper is organized as follows. In Sec. \ref{theory}, we present the theoretical framework, including the derivation of the effective potentials for the $\Xi_b^{(\prime,\,*)}B^{(*)}$ systems and the $G$-parity rule that connects them to their $\Xi_b^{(\prime,\,*)}\bar B^{(*)}$ counterparts. In Sec. \ref{results}, we present the bound-state properties of both the $\Xi_b^{(\prime,\,*)}B^{(*)}$ and $\Xi_b^{(\prime,\,*)}\bar B^{(*)}$ systems, and predict the promising candidates of the singly-strange hidden-bottom and double-bottom molecular pentaquarks. A brief summary and outlook are given in Sec. \ref{summary}.

\section{Theoretical Framework}\label{theory}

In this section, we first present the derivation of the coordinate-space effective potentials for the $\Xi_b^{(\prime,\,*)}B^{(*)}$ systems within the OBE model. The derivation consists of three successive stages. We begin by constructing the effective Lagrangians that govern the couplings of the heavy hadrons to the light scalar ($\sigma$), pseudoscalar ($\pi$,\,$\eta$), and vector ($\rho$,\,$\omega$) mesons. The masses of these exchanged mesons are taken as $m_\sigma = 600.00$ MeV, $m_\pi = 137.27$ MeV, $m_\eta = 547.86$ MeV, $m_\rho = 775.26$ MeV, and $m_\omega = 782.66$ MeV \cite{ParticleDataGroup:2024cfk}. From these effective Lagrangians, we then compute the scattering amplitudes for the relevant two-body scattering processes. Finally, using the Breit approximation \cite{Berestetskii:1982qgu} followed by a Fourier transformation, we convert these scattering amplitudes into the corresponding coordinate-space effective potentials. For the $\Xi_b^{(\prime,\,*)}\bar B^{(*)}$ systems, the corresponding effective potentials are obtained from the $\Xi_b^{(\prime,\,*)}B^{(*)}$ counterparts via the $G$-parity rule \cite{Klempt:2002ap}, which establish a clear and consistent link between the particle and antiparticle channels.

\subsection{Effective Lagrangians}

Our investigation of the mass spectra of the singly-strange hidden-bottom molecular pentaquark states is built upon the OBE model, which provides a systematic framework for describing the effective interactions among the constituent hadrons via the exchange of the light mesons. The constituent hadrons in our study are the $B^{(*)}$ mesons and the bottom-strange baryons $\Xi_b^{(\prime,\,*)}$, with their masses taken from the Particle Data Group \cite{ParticleDataGroup:2024cfk}: $m_B = 5279.57$ MeV, $m_{B^*} = 5324.75$ MeV, $m_{\Xi_b} = 5794.35$ MeV, $m_{\Xi_b^\prime} = 5934.90$ MeV, and $m_{\Xi_b^*} = 5953.90$ MeV. In order to construct the interaction vertices entering the OBE effective potentials, we employ the effective Lagrangian formalism that is constrained by the HQSS, the chiral symmetry, and the hidden local symmetry (HLS) \cite{Wise:1992hn,Casalbuoni:1992gi,Casalbuoni:1996pg,Yan:1992gz,Harada:2003jx,Bando:1987br}.

We begin with the single heavy meson sector. In the heavy-quark limit \cite{Wise:1992hn}, the spin symmetry implies that the pseudoscalar meson $B$ and the vector meson $B^*$ are degenerate and thus form a spin doublet. It proves convenient to assemble these states into a single heavy meson superfield $H^{(\overline{Q})}_a$ \cite{Ding:2008gr}:
\begin{equation}
H^{(\overline{Q})}_a=\left(\widetilde{\mathcal{P}}^{*\mu}_{a}\gamma_{\mu}-\widetilde{\mathcal{P}}_a\gamma_5\right)\frac{1-\slash \!\!\!v}{2},
\end{equation}
where the flavor index $a=u,\,d,\,s$ labels the light quark content. The fields $\widetilde{\mathcal{P}}$ and $\widetilde{\mathcal{P}}^*$ represent the pseudoscalar and vector heavy mesons, respectively, with $\widetilde{\mathcal{P}} = \left(B^{+},\,B^{0},\,B_s^{0} \right)$ and $\widetilde{\mathcal{P}}^{*} = \left(B^{*+},\,B^{*0},\,B_s^{*0} \right)$. The four-velocity of the heavy meson is denoted by $v^\mu=(1,\,\mathbf{0})$, and the conjugate field follows the standard definition $\bar H^{(\overline{Q})}_a = \gamma^0 H^{(\overline{Q})\dagger}_a \gamma^0$ \cite{Ding:2008gr}.

Imposing the simultaneous constraints of HQSS, chiral symmetry, and HLS, the interaction Lagrangians describing the couplings of the single heavy meson superfield to the light mesons take the form \cite{Ding:2008gr}:
\begin{eqnarray}
\mathcal{L}_{HH\mathcal{E}}&=&g_S\langle \bar{H}^{(\overline{Q})}_a\sigma H^{(\overline{Q})}_b\rangle+ig\langle \bar{H}^{(\overline{Q})}_a\gamma_{\mu}{\mathcal A}_{ab}^{\mu}\gamma_5H^{(\overline{Q})}_b\rangle\nonumber\\
  &&-i\beta\langle \bar{H}^{(\overline{Q})}_av_{\mu}\left(\mathcal{V}^{\mu}-\rho^{\mu}\right)_{ab}H^{(\overline{Q})}_b\rangle\nonumber\\
  &&+i\lambda\langle \bar{H}^{(\overline{Q})}_a\sigma_{\mu\nu}F^{\mu\nu}_{ab}(\rho)H^{(\overline{Q})}_b\rangle.
\end{eqnarray}
Here, $\mathcal{E}$ denotes the exchanged light meson. The first term in $\mathcal L_{HH\mathcal E}$ describes the scalar-meson exchange, the second term governs the pseudoscalar-meson couplings, while the last two terms account for the vector-meson interactions. $\mathcal{A}_\mu$ and ${\mathcal V}_{\mu}$ are the axial and vector currents built from the chiral field $\xi=e^{i\mathbb{P}/f_\pi}$:
\begin{eqnarray}
{\mathcal A}_{\mu}=\frac{1}{2}\left(\xi^{\dagger}\partial_{\mu}\xi-\xi\partial_{\mu}\xi^{\dagger}\right)~{\rm and}~
{\mathcal V}_{\mu}=\frac{1}{2}\left(\xi^{\dagger}\partial_{\mu}\xi+\xi\partial_{\mu}\xi^{\dagger}\right),
\end{eqnarray}
where $f_\pi$ is the pion decay constant, and $\mathbb{P}$ denotes the pseudoscalar meson matrix \cite{Wise:1992hn,Casalbuoni:1992gi,Casalbuoni:1996pg,Yan:1992gz,Bando:1987br,Harada:2003jx,Chen:2017xat,Ding:2008gr}:
\begin{eqnarray*}
{\mathbb{P}} &=& {\left(\begin{array}{ccc}
       \frac{\pi^0}{\sqrt{2}}+\frac{\eta}{\sqrt{6}} &\pi^+ &K^+\\
       \pi^-       &-\frac{\pi^0}{\sqrt{2}}+\frac{\eta}{\sqrt{6}} &K^0\\
       K^-         &\bar K^0   &-\sqrt{\frac{2}{3}} \eta     \end{array}\right)}.
\end{eqnarray*}
The light vector mesons enter through the field $\rho_{\mu}=i{g_V}\mathbb{V}_{\mu}/{\sqrt{2}}$ and the corresponding field strength tensor $F_{\mu\nu}(\rho)=\partial_{\mu}\rho_{\nu}-\partial_{\nu}\rho_{\mu}+[\rho_{\mu},\,\rho_{\nu}]$, with the vector meson matrix $\mathbb{V}_{\mu}$ given by \cite{Wise:1992hn,Casalbuoni:1992gi,Casalbuoni:1996pg,Yan:1992gz,Bando:1987br,Harada:2003jx,Chen:2017xat,Ding:2008gr}
\begin{eqnarray*}
{\mathbb{V}}_{\mu} &=& {\left(\begin{array}{ccc}
       \frac{\rho^0}{\sqrt{2}}+\frac{\omega}{\sqrt{2}} &\rho^+ &K^{*+}\\
       \rho^-       &-\frac{\rho^0}{\sqrt{2}}+\frac{\omega}{\sqrt{2}} &K^{*0}\\
       K^{*-}         &\bar K^{*0}   & \phi     \end{array}\right)}_{\mu}.
\end{eqnarray*}

Turning to the single heavy baryon sector, the $S$-wave bottom-strange baryons with $J^P=1/2^+$ appear in two distinct SU(3) flavor representations: $\Xi_b$ resides in the anti-triplet $\bar{3}_F$, whereas $\Xi_b^\prime$ belongs to the sextet $6_F$. The spin-$3/2$ partner $\Xi_b^*$ ($J^P=3/2^+$) is grouped with $\Xi_b^\prime$ under HQSS. Following the standard single heavy baryon formalism, the sextet single heavy baryons $\mathcal{B}_6$ and $\mathcal{B}_6^*$ are unified into a superfield $\mathcal{S}_{\mu}$ defined by \cite{Wise:1992hn,Casalbuoni:1992gi,Casalbuoni:1996pg,Yan:1992gz,Bando:1987br,Harada:2003jx,Chen:2017xat}
\begin{eqnarray}
\mathcal{S}_{\mu}=-\sqrt{\frac{1}{3}}(\gamma_{\mu}+v_{\mu})\gamma^5\mathcal{B}_6+\mathcal{B}_{6\mu}^*,
\end{eqnarray}
where $\bar{\mathcal{S}}_{\mu}=\mathcal{S}_{\mu}^\dag\gamma^0$ is the Dirac conjugate. The explicit matrix representations of the anti-triplet and sextet bottom baryon fields take the standard forms \cite{Wise:1992hn,Casalbuoni:1992gi,Casalbuoni:1996pg,Yan:1992gz,Bando:1987br,Harada:2003jx,Chen:2017xat}
\begin{eqnarray*}
\mathcal{B}_{\bar{3}} &=& \left(\begin{array}{ccc}
        0    &\Lambda_b^0      &\Xi_b^0\\
        -\Lambda_b^0  &0      &\Xi_b^-\\
        -\Xi_b^0      &-\Xi_b^-     &0
\end{array}\right),\\
\mathcal{B}_6^{(*)} &=& \left(\begin{array}{ccc}
         \Sigma_b^{{(*)}+}                  &\frac{\Sigma_b^{{(*)}0}}{\sqrt{2}}     &\frac{\Xi_b^{\prime(*)0}}{\sqrt{2}}\\[1mm]
         \frac{\Sigma_b^{{(*)}0}}{\sqrt{2}}      &\Sigma_b^{{(*)}-}    &\frac{\Xi_b^{\prime(*)-}}{\sqrt{2}}\\[1mm]
         \frac{\Xi_b^{\prime(*)0}}{\sqrt{2}}    &\frac{\Xi_b^{\prime(*)-}}{\sqrt{2}}      &\Omega_b^{(*)-}
\end{array}\right).
\end{eqnarray*}

Guided by the same symmetry principles, i.e., HQSS, chiral symmetry, and HLS, the effective Lagrangians governing the interactions between the bottom baryons and the light mesons are constructed as follows \cite{Wise:1992hn,Casalbuoni:1992gi,Casalbuoni:1996pg,Yan:1992gz,Bando:1987br,Harada:2003jx,Chen:2017xat}:
\begin{eqnarray}
\mathcal{L}_{\mathcal{B}_{\bar{3}}\mathcal{B}_{\bar{3}}\mathcal{E}} &=& l_B\langle\bar{\mathcal{B}}_{\bar{3}}\sigma\mathcal{B}_{\bar{3}}\rangle
          +i\beta_B\langle\bar{\mathcal{B}}_{\bar{3}}v^{\mu}(\mathcal{V}_{\mu}
          -\rho_{\mu})\mathcal{B}_{\bar{3}}\rangle,\\[1mm]
\mathcal{L}_{\mathcal{S}\mathcal{S}\mathcal{E}} &=& l_S\langle\bar{\mathcal{S}}_{\mu}\sigma\mathcal{S}^{\mu}\rangle -\frac{3}{2}g_1\varepsilon^{\mu\nu\lambda\kappa}v_{\kappa}\langle\bar{\mathcal{S}}_{\mu}\mathcal{A}_{\nu}
\mathcal{S}_{\lambda}\rangle\nonumber\\
    &&+i\beta_{S}\langle\bar{\mathcal{S}}_{\mu}v_{\alpha}\left(\mathcal{V}^{\alpha}
    -\rho^{\alpha}\right) \mathcal{S}^{\mu}\rangle +\lambda_S\langle\bar{\mathcal{S}}_{\mu}F^{\mu\nu}(\rho)\mathcal{S}_{\nu}\rangle,\nonumber\\\\[1mm]
\mathcal{L}_{\mathcal{B}_{\bar{3}}\mathcal{S}\mathcal{E}}&=&ig_4\langle\bar{\mathcal{S}^{\mu}}
\mathcal{A}_{\mu}\mathcal{B}_{\bar{3}}\rangle+i\lambda_I\varepsilon^{\mu\nu\lambda
\kappa}v_{\mu}\langle \bar{\mathcal{S}}_{\nu}F_{\lambda\kappa}(\rho)\mathcal{B}_{\bar{3}}\rangle+h.c..\nonumber\\
\end{eqnarray}
In the above effective Lagrangians, the terms proportional to $\sigma$ describe the scalar-meson exchanges, those involving $\mathcal{A}_\mu$ correspond to the pseudoscalar-meson couplings, while those containing $\mathcal V_\mu$, $\rho_\mu$, and $F_{\mu\nu}(\rho)$ arise from the vector-meson interactions. The coupling constants $l_B$, $l_S$, $g_1$, $g_4$, $\beta_B$, $\beta_S$, $\lambda_S$, and $\lambda_I$ encode the strengths of the interaction vertices.

In the following, we expand the compact effective Lagrangians constructed above to leading order in  the chiral field. This expansion yields the explicit interaction vertices that are directly employed in the evaluation of the scattering amplitudes for the relevant two-body scattering processes. For clarity, we present the resulting effective Lagrangians according to the type of exchanged meson. The expanded effective Lagrangians for the scalar-meson couplings are:
\begin{eqnarray}
\mathcal{L}_{\widetilde{\mathcal{P}}^{(*)}\widetilde{\mathcal{P}}^{(*)}\sigma} &=&-2g_S{\widetilde{\mathcal{P}}}_a {\widetilde{\mathcal{P}}}_a^{\dag} \sigma+ 2g_S {\widetilde{\mathcal{P}}}_{a\mu}^* {\widetilde{\mathcal{P}}}_a^{*\mu\dag} \sigma,\\[1mm]
\mathcal{L}_{\mathcal{B}_{\bar{3}}\mathcal{B}_{\bar{3}}\sigma}&=& l_B\langle \bar{\mathcal{B}}_{\bar{3}}\sigma\mathcal{B}_{\bar{3}}\rangle,\\[1mm]
\mathcal{L}_{\mathcal{B}_{6}^{(*)}\mathcal{B}_{6}^{(*)}\sigma}&=&-l_S\langle\bar{\mathcal{B}}_6\sigma\mathcal{B}_6\rangle+l_S\langle
\bar{\mathcal{B}}_{6\mu}^{*}
\sigma\mathcal{B}_6^{*\mu}\rangle\nonumber\\
    &&-\frac{l_S}{\sqrt{3}}\langle\bar{\mathcal{B}}_{6\mu}^{*}\sigma
    \left(\gamma^{\mu}+v^{\mu}\right)\gamma^5\mathcal{B}_6\rangle+h.c.,
\end{eqnarray}
and we have the following pseudoscalar-meson interaction Lagrangians:
\begin{eqnarray}
\mathcal{L}_{\widetilde{\mathcal{P}}^{(*)}\widetilde{\mathcal{P}}^{(*)}\mathbb{P}}&=&\frac{2ig}{f_{\pi}}v^{\alpha}\varepsilon_{\alpha\mu\nu\lambda}{\widetilde{\mathcal{P}}}_a^{*\mu\dag}{\widetilde{\mathcal{P}}}_b^{*\lambda}\partial^{\nu}{\mathbb{P}}_{ab}\nonumber\\
    &&+\frac{2g}{f_{\pi}}\left({\widetilde{\mathcal{P}}}_a^{*\mu\dag}{\widetilde{\mathcal{P}}}_b+{\widetilde{\mathcal{P}}}_a^{\dag}{\widetilde{\mathcal{P}}}_b^{*\mu}\right)\partial_{\mu}{\mathbb{P}}_{ab},\\[1mm]
\mathcal{L}_{\mathcal{B}_6^{(*)}\mathcal{B}_6^{(*)}\mathbb{P}}&=&i\frac{g_1}{2f_{\pi}}\varepsilon^{\mu\nu\lambda\kappa}v_{\kappa}\langle\bar{\mathcal{B}}_6
\gamma_{\mu}\gamma_{\lambda}\partial_{\nu}\mathbb{P}\mathcal{B}_6\rangle\nonumber\\
    &&-i\frac{3g_1}{2f_{\pi}}\varepsilon^{\mu\nu\lambda\kappa}v_{\kappa}
    \langle\bar{\mathcal{B}}_{6\mu}^{*}\partial_{\nu}\mathbb{P}\mathcal{B}_{6\lambda}^*\rangle\nonumber\\
    &&+i\frac{\sqrt{3}g_1}{2f_{\pi}}v_{\kappa}\varepsilon^{\mu\nu\lambda\kappa}
    \langle\bar{\mathcal{B}}_{6\mu}^*\partial_{\nu}\mathbb{P}{\gamma_{\lambda}\gamma^5}
      \mathcal{B}_6\rangle+h.c.,\nonumber\\\\[1mm]
\mathcal{L}_{\mathcal{B}_{\bar{3}}\mathcal{B}_6^{(*)}\mathbb{P}} &=& -\sqrt{\frac{1}{3}}\frac{g_4}{f_{\pi}}\langle\bar{\mathcal{B}}_6\gamma^5\left(\gamma^{\mu}
+v^{\mu}\right)\partial_{\mu}\mathbb{P}\mathcal{B}_{\bar{3}}\rangle\nonumber\\
    &&-\frac{g_4}{f_{\pi}}\langle\bar{\mathcal{B}}_{6\mu}^*\partial^{\mu} \mathbb{P}\mathcal{B}_{\bar{3}}\rangle+h.c.,
\end{eqnarray}
while the vector-meson interaction Lagrangians are given by
\begin{eqnarray}
\mathcal{L}_{{\widetilde{\mathcal{P}}}^{(*)}{\widetilde{\mathcal{P}}}^{(*)}\mathbb{V}} &=&\sqrt{2}\beta g_V {\widetilde{\mathcal{P}}}_a {\widetilde{\mathcal{P}}}_b^{\dag} v\cdot\mathbb{V}_{ab}-\sqrt{2}\beta g_V {\widetilde{\mathcal{P}}}_{a\mu}^* {\widetilde{\mathcal{P}}}_b^{*\mu\dag}v\cdot\mathbb{V}_{ab}\nonumber\\
    &&-2\sqrt{2}i\lambda g_V {\widetilde{\mathcal{P}}}_a^{*\mu\dag}{\widetilde{\mathcal{P}}}_b^{*\nu}\left(\partial_{\mu}\mathbb{V}_{\nu}-\partial_{\nu}\mathbb{V}_{\mu}\right)_{ab}\nonumber\\
    &&-2\sqrt{2}\lambda g_V v^{\lambda}\varepsilon_{\lambda\mu\alpha\beta}\left({\widetilde{\mathcal{P}}}_a^{*\mu\dag}{\widetilde{\mathcal{P}}}_b+{\widetilde{\mathcal{P}}}_a^{\dag}{\widetilde{\mathcal{P}}}_b^{*\mu}\right)\partial^{\alpha}\mathbb{V}^{\beta}_{ab},\nonumber\\\\[1mm]
\mathcal{L}_{\mathcal{B}_{\bar{3}}\mathcal{B}_{\bar{3}}\mathbb{V}}&=&
\frac{1}{\sqrt{2}}\beta_Bg_V\langle\bar{\mathcal{B}}_{\bar{3}}v\cdot\mathbb{V}
\mathcal{B}_{\bar{3}}\rangle,\\[1mm]
\mathcal{L}_{\mathcal{B}_6^{(*)}\mathcal{B}_6^{(*)}\mathbb{V}}&=&
-\frac{\beta_Sg_V}{\sqrt{2}}\langle\bar{\mathcal{B}}_6v\cdot\mathbb{V}
\mathcal{B}_6\rangle+\frac{\beta_Sg_V}{\sqrt{2}}\langle\bar{\mathcal{B}}_{6\mu}^*v\cdot {V}\mathcal{B}_6^{*\mu}\rangle\nonumber\\
    &&-i\frac{\lambda_S g_V}{3\sqrt{2}}\langle\bar{\mathcal{B}}_6\gamma_{\mu}\gamma_{\nu}
    \left(\partial^{\mu}\mathbb{V}^{\nu}-\partial^{\nu}\mathbb{V}^{\mu}\right)
    \mathcal{B}_6\rangle\nonumber\\
    &&+i\frac{\lambda_Sg_V}{\sqrt{2}}\langle\bar{\mathcal{B}}_{6\mu}^*
    \left(\partial^{\mu}\mathbb{V}^{\nu}-\partial^{\nu}\mathbb{V}^{\mu}\right)
    \mathcal{B}_{6\nu}^*\rangle\nonumber\\
    &&-i\frac{\lambda_Sg_V}{\sqrt{6}}\langle\bar{\mathcal{B}}_{6\mu}^*
    \left(\partial^{\mu}\mathbb{V}^{\nu}-\partial^{\nu}\mathbb{V}^{\mu}\right)
    \left(\gamma_{\nu}+v_{\nu}\right)\gamma^5\mathcal{B}_6\rangle\nonumber\\
    &&-\frac{\beta_Sg_V}{\sqrt{6}}\langle\bar{\mathcal{B}}_{6\mu}^*v\cdot \mathbb{V}\left(\gamma^{\mu}+v^{\mu}\right)\gamma^5\mathcal{B}_6\rangle+h.c.,\\[1mm]
\mathcal{L}_{\mathcal{B}_{\bar{3}}\mathcal{B}_6^{(*)}\mathbb{V}} &=&
-\frac{\lambda_Ig_V}{\sqrt{2}}\varepsilon^{\mu\nu\lambda\kappa}v_{\mu}\langle \bar{\mathcal{B}}_{6\nu}^*\left(\partial_{\lambda}\mathbb{V}_{\kappa}
-\partial_{\kappa}\mathbb{V}_{\lambda}\right)
          \mathcal{B}_{\bar{3}}\rangle\nonumber\\
    &&-\frac{\lambda_Ig_V}{\sqrt{6}}\varepsilon^{\mu\nu\lambda\kappa}v_{\mu}\langle \bar{\mathcal{B}}_6\gamma^5\gamma_{\nu}
        \left(\partial_{\lambda}\mathbb{V}_{\kappa}-\partial_{\kappa} \mathbb{V}_{\lambda}\right)\mathcal{B}_{\bar{3}}\rangle+h.c..\nonumber\\
\end{eqnarray}

To perform quantitative calculations based on the effective Lagrangians derived above, one must assign definite values to the coupling constants appearing in the interaction vertices. These coupling constants are determined by a combination of experimental constraints and theoretical estimates, including the quark-model computations, the QCD sum-rule analyses, the lattice-QCD simulations, and so on. According to the type of exchanged meson, they can be grouped into the scalar, pseudoscalar, and vector couplings. The full set of coupling constants adopted in this work is compiled in the following:
\begin{center}
\renewcommand{\arraystretch}{1.50}
\begin{tabular*}{85mm}{@{\extracolsep{\fill}}lll}
\toprule[1pt]
\toprule[1pt]
Categories & Couplings & Values \\
\hline
\multirow{2}{*}{Scalar}
  & $g_S$ & $0.76$ \\
  & $l_B$, $l_S$ & $-3.65$, $6.20$ \\
\hline
\multirow{3}{*}{Pseudoscalar}
  & $g$ & $0.59$ \\
  & $g_1$, $g_4$ & $0.94$, $1.06$\\
  & $f_\pi$ & $132$ MeV \\
\hline
\multirow{2}{*}{Vector}
  & $\beta g_V$, $\beta_B g_V$, $\beta_S g_V$ & $-5.25$, $-6.00$, $12.00$ \\
  & $\lambda g_V$, $\lambda_S g_V$, $\lambda_I g_V$ & $-3.27$, $19.20$, $-6.80$~GeV$^{-1}$ \\
\bottomrule[1pt]\bottomrule[1pt]
\end{tabular*}.
\end{center}
Here, we need to mention that their values are consistent with those commonly used in  Refs. \cite{Wang:2022mxy,Chen:2017xat,Chen:2019asm,Chen:2020kco,Wang:2020bjt,Wang:2019nwt,Chen:2018pzd,Wang:2021hql,Wang:2021yld,Wang:2021aql,Yang:2021sue,Wang:2020dya,Wang:2023ftp,Yalikun:2023waw,Wang:2023aob,Wang:2023ael,Wang:2026prg}.

\subsection{Effective potentials for the $\Xi_b^{(\prime,\,*)}B^{(*)}$ and $\Xi_b^{(\prime,\,*)}\bar B^{(*)}$ systems}

Before proceeding to the derivation of the effective potentials, we first set up the isospin and spin-orbital wave functions for the $\Xi_b^{(\prime,\,*)} B^{(*)}$ systems. Since both the bottom-strange baryon $\Xi_b^{(\prime,\,*)}$ and the bottom meson $B^{(*)}$ carry isospin $I=1/2$, their flavor structures need to be specified in order to evaluate the matrix elements of the OBE effective potentials. The $\Xi_b^{(\prime,\,*)}$ baryon exists as an isospin doublet,
\begin{eqnarray*}
|\Xi_b^{(\prime,*)0}\rangle = \Big|I=\frac12, I_3=+\frac12\Big\rangle, ~~~~~
|\Xi_b^{(\prime,*)-}\rangle = \Big|I=\frac12, I_3=-\frac12\Big\rangle,
\end{eqnarray*}
and the $B^{(*)}$ meson forms the corresponding isospin doublet,
\begin{eqnarray*}
|B^{(*)+}\rangle = \Big|I=\frac12, I_3=+\frac12\Big\rangle, ~~~~~
|B^{(*)0}\rangle = \Big|I=\frac12, I_3=-\frac12\Big\rangle.
\end{eqnarray*}
By coupling two isospin-$1/2$ states via the Clebsch-Gordan coefficients, one obtains the isospin wave functions for the $\Xi_b^{(\prime,\,*)} B^{(*)}$ systems:
\begin{eqnarray}
\Big|I=0, I_3=0\Big\rangle &=& \frac{1}{\sqrt{2}}|\Xi_b^{(\prime,*)0} B^{(*)0}\rangle-\frac{1}{\sqrt{2}} |\Xi_b^{(\prime,*)-} B^{(*)+}\rangle,\nonumber\\\\[1mm]
\Big|I=1, I_3=+1\Big\rangle &=& |\Xi_b^{(\prime,*)0} B^{(*)+}\rangle,\\[1mm]
\Big|I=1, I_3=0\Big\rangle &=& \frac{1}{\sqrt{2}}|\Xi_b^{(\prime,*)0} B^{(*)0}\rangle+\frac{1}{\sqrt{2}} |\Xi_b^{(\prime,*)-} B^{(*)+}\rangle,\nonumber\\\\[1mm]
\Big|I=1, I_3=-1\Big\rangle &=& |\Xi_b^{(\prime,*)-} B^{(*)0}\rangle.
\end{eqnarray}
These isospin configurations will be used in the evaluation of the OBE potential matrix elements, as the effective interactions depend on the total isospin through the exchange of the light mesons, notably the $\pi$ and $\rho$.

For the spin and orbital part, we adopt the spectroscopic notation $|{}^{2S+1}L_J\rangle$, where the total spin $\bm{S}=\bm{S}_1+\bm{S}_2$ arises from coupling the bottom-strange baryon spin $\bm{S}_1$ and the bottom meson spin $\bm{S}_2$, while the total angular momentum is $\bm{J}=\bm{S}+\bm{L}$. The explicit spin and orbital wave function reads
\begin{equation}
\big|{}^{2S+1}L_J\rangle = \sum_{m_S,\,m_L}C^{JM}_{S m_S,\,L m_L}\,\big|S m_S\big\rangle\,\big|Y_{L m_L}\big\rangle,
\end{equation}
where $C^{JM}_{S m_S,\,L m_L}$ denotes the Clebsch-Gordan coefficient, and $|Y_{L m_L}\rangle$ is the spherical harmonic describing the relative orbital motion. In the present work, we focus on the $S$-wave as the leading partial wave that provides the main binding drive, and the $D$-wave is retained only when the tensor force induces the $S$-$D$ wave mixing effects. The spin-parity quantum numbers and the accessible channels for each $\Xi_b^{(\prime,\,*)}B^{(*)}$ system are summarized as follows:
\begin{center}
\renewcommand{\arraystretch}{1.50}
\begin{tabular*}{86mm}{@{\extracolsep{\fill}}ccl}
\toprule[1pt]
\toprule[1pt]
Systems & $J^P$ & $|{}^{2S+1}L_J\rangle$ ($L=0,\,2$) \\
\hline
$\Xi_b B$,\,$\Xi_b' B$ & $\frac{1}{2}^-$ & $|{}^2S_{1/2}\rangle$\\
\hline
\multirow{2}{*}{$\Xi_b B^*$,\,$\Xi_b' B^*$}
 & $\frac{1}{2}^-$ & $|{}^2S_{1/2}\rangle$, $|{}^4D_{1/2}\rangle$ \\[1mm]
 & $\frac{3}{2}^-$ & $|{}^4S_{3/2}\rangle$, $|{}^2D_{3/2}\rangle$, $|{}^4D_{3/2}\rangle$\\
\hline
$\Xi_b^* B$ & $\frac{3}{2}^-$ & $|{}^4S_{3/2}\rangle$, $|{}^4D_{3/2}\rangle$\\
\hline
\multirow{3}{*}{$\Xi_b^* B^*$}
 & $\frac12^-$ & $|{}^2S_{1/2}\rangle$, $|{}^4D_{1/2}\rangle$, $|{}^6D_{1/2}\rangle$ \\[1mm]
 & $\frac32^-$ & $|{}^4S_{3/2}\rangle$, $|{}^2D_{3/2}\rangle$, $|{}^4D_{3/2}\rangle$, $|{}^6D_{3/2}\rangle$ \\[1mm]
 & $\frac52^-$ & $|{}^6S_{5/2}\rangle$, $|{}^2D_{5/2}\rangle$, $|{}^4D_{5/2}\rangle$, $|{}^6D_{5/2}\rangle$ \\
\bottomrule[1pt]\bottomrule[1pt]
\end{tabular*}.
\end{center}

Having established the isospin and spin-orbital wave functions, we now deduce the coordinate-space effective potentials for the $\Xi_b^{(\prime,\,*)}B^{(*)}$ systems. Starting from the effective Lagrangians given above, we compute the scattering amplitude $\mathcal{M}_{\mathcal{E}}^{\,\Xi_b^{(\prime,\,*)} B^{(*)}\to\Xi_b^{(\prime,\,*)} B^{(*)}}(\bm{q})$ for the process $\Xi_b^{(\prime,\,*)} B^{(*)} \to \Xi_b^{(\prime,\,*)} B^{(*)}$ with the exchange of a light meson $\mathcal{E}$. And then, the scattering amplitude is related to the momentum-space effective potential $\mathcal{V}_{\mathcal{E}}^{\,\Xi_b^{(\prime,\,*)} B^{(*)}\to\Xi_b^{(\prime,\,*)} B^{(*)}}(\bm{q})$ through the Breit approximation \cite{Berestetskii:1982qgu}:
\begin{equation}
\mathcal{V}_{\mathcal{E}}^{\,\Xi_b^{(\prime,\,*)} B^{(*)}\to\Xi_b^{(\prime,\,*)} B^{(*)}}(\bm{q})
= -\frac{\mathcal{M}_{\mathcal{E}}^{\,\Xi_b^{(\prime,\,*)} B^{(*)}\to\Xi_b^{(\prime,\,*)} B^{(*)}}(\bm{q})}
{4\sqrt{m_{\Xi_b^{(\prime,\,*)}}\,m_{B^{(*)}}m_{\Xi_b^{(\prime,\,*)}}\,m_{B^{(*)}}}},
\end{equation}
where $\bm{q}$ is the three-momentum carried by the exchanged meson. Since the interaction vertices are of finite spatial extent, we introduce the form factor $\mathcal{F}(q^2) = \Lambda^2/(\Lambda^2-q^2)$ at each interaction vertex, where $\Lambda$ is the cutoff parameter. The same form factor has been widely adopted in studies of the hadronic molecules \cite{He:2015mja,Lu:2016nlp,Chen:2017vai,Ding:2021igr,Wang:2021ajy,Ding:2023yuo,Lin:2024qcq,Ding:2024dif,Qian:2024joy,Wang:2024ukc,Chen:2025gxe,Wang:2025kpm,Cheng:2026cgo,Feng:2026wbl}. It is noteworthy that the binding properties of the deuteron as a loosely bound state composed of a proton and a neutron can be reproduced with $\Lambda \simeq 0.38$ GeV within the OBE framework. Details on the effective Lagrangians and the coupling constants can be found in Ref. \cite{Chen:2017jjn}. Following this established treatment, we regard a loosely bound state obtained with $\Lambda$ around $0.38$ GeV as a strong indication of a hadronic molecular state. The coordinate-space effective potential $\mathcal{V}_{\mathcal{E}}^{\,\Xi_b^{(\prime,\,*)} B^{(*)}\to\Xi_b^{(\prime,\,*)} B^{(*)}}(\bm{r})$ can be obtained via the Fourier transform of the regulated momentum-space effective potential:
\begin{equation}
\mathcal{V}_{\mathcal{E}}^{\,\Xi_b^{(\prime,\,*)} B^{(*)}\to\Xi_b^{(\prime,\,*)} B^{(*)}}(\bm{r}) = \int\!\frac{d^3qe^{i\bm{q}\cdot\bm{r}}}{(2\pi)^3}\mathcal{V}_{\mathcal{E}}^{\,\Xi_b^{(\prime,\,*)} B^{(*)}\to\Xi_b^{(\prime,\,*)} B^{(*)}}(\bm{q})\mathcal{F}^2(q^2).
\end{equation}

After performing the integration and collecting the spin structures, the obtained effective potential decomposes into a central part, a spin-spin term, and a tensor piece:
\begin{equation}
\mathcal{V}(r) = \mathcal{V}^C(r)+\mathcal{V}^{SS}(r)\bm{S}_1\cdot\bm{S}_2
+ \mathcal{V}^T(r)\,S_{12}(\bm{S}_1, \bm{S}_2, \hat{\bm{r}}),
\end{equation}
where $S_{12}(\bm{S}_1, \bm{S}_2, \hat{\bm{r}})=3(\bm{S}_1\cdot\hat{\bm{r}})
(\bm{S}_2\cdot\hat{\bm{r}})
-\bm{S}_1\cdot\bm{S}_2$ is the tensor operator. The  expressions of $\mathcal{V}^C(r)$, $\mathcal{V}^{SS}(r)$, and $\mathcal{V}^T(r)$ for the $\Xi_b^{(\prime,\,*)} B^{(*)}$ systems can refer to Refs. \cite{Chen:2020kco, Wang:2022mxy, Chen:2022onm}. They are built from the linear combinations of the Yukawa-type functions regulated by the form factor, with the coefficients fixed by the coupling constants and the isospin factors.

The effective potentials for the $\Xi_b^{(\prime,\,*)}\bar B^{(*)}$ double-bottom systems can be efficiently obtained from the $\Xi_b^{(\prime,\,*)}B^{(*)}$ hidden-bottom systems through the $G$-parity rule \cite{Klempt:2002ap}. For the $\Xi_b^{(\prime,\,*)} \bar B^{(*)}\to\Xi_b^{(\prime,\,*)} \bar B^{(*)}$ processes, the corresponding scattering amplitudes transform as:
\begin{equation}
\mathcal{M}_{\mathcal{E}}^{\,\Xi_b^{(\prime,\,*)} \bar B^{(*)}\to\Xi_b^{(\prime,\,*)} \bar B^{(*)}}(\bm{q}) = G_{\mathcal{E}} \cdot \mathcal{M}_{\mathcal{E}}^{\,\Xi_b^{(\prime,\,*)} B^{(*)}\to\Xi_b^{(\prime,\,*)} B^{(*)}}(\bm{q}),
\end{equation}
where $G_{\mathcal{E}}$ is the $G$-parity of the exchanged meson: $G_\sigma = +1$, $G_\pi = -1$, $G_\eta = +1$, $G_\rho = +1$, and $G_\omega = -1$ \cite{ParticleDataGroup:2024cfk}. This powerful relation allows us to obtain the $\Xi_b^{(\prime,\,*)}\bar B^{(*)}$ effective potentials directly from the previously derived $\Xi_b^{(\prime,\,*)}B^{(*)}$ effective potentials without recalculating the corresponding scattering amplitudes.

\section{Numerical results and discussions}\label{results}

In this section, we discuss the bound-state properties for the $\Xi_b^{(\prime,\,*)}B^{(*)}$ and $\Xi_b^{(\prime,\,*)}\bar B^{(*)}$ systems, and predict the promising candidates of the singly-strange hidden-bottom and double-bottom molecular pentaquarks. We systematically solve the coupled-channel Schr\"odinger equation for all relevant isospin and spin-parity quantum numbers taking into account both the single-channel and coupled-channel cases. For each identified bound state, we provide the binding energy $E$, the root-mean-square (RMS) radius $r_{\rm RMS}$, and the probabilities of the various components $P(\alpha)$. To illustrate the cutoff dependence of the obtained bound-state properties, we provide three representative sets of solutions for each loosely bound state, corresponding to different values with the cutoff parameter $\Lambda$ chosen within the reasonable range. For the isoscalar $\Xi_b^{(\prime,\,*)}B^{(*)}$ and $\Xi_b^{(\prime,\,*)}\bar B^{(*)}$ systems, we find the existent of the loosely bound-state solutions across all systems considered. In contrast, for the isovector $\Xi_b^{(\prime,\,*)}B^{(*)}$ and $\Xi_b^{(\prime,\,*)}\bar B^{(*)}$ systems, no loosely bound-state solutions are obtained for the cutoff parameter $\Lambda$ within the reasonable range. This clear distinction implies that the effective interaction in the isovector system is not sufficiently attractive to bind the constituent hadrons, whereas the isoscalar system provides the required binding strength to form the loosely bound state.

\subsection{Singly-strange hidden-bottom $\Xi_b^{(\prime,\,*)} B^{(*)}$ systems}

\paragraph{$\Xi_b B$ system}

We first discuss the bound-state properties for the $\Xi_b B$ state with quantum numbers $I(J^P)=0(1/2^-)$. As shown in Table \ref{XibB1},  we can obtain the loosely bound-state solutions by confining the single-channel treatment to the $\Xi_b B({}^2S_{1/2})$ channel. Taking three representative cutoff parameters $\Lambda = 0.35$, $0.45$, and 0.55 GeV, the corresponding binding energies are found to be $E = -2.08$, $-5.86$, and $-11.84$ MeV, while the associated RMS radii shrink progressively from 2.00 fm down to 0.99 fm. This trend is consistent with the expected behavior: a larger cutoff enhances the effective attraction and compresses the spatial wave function, leading to the hadronic configuration with deeper binding and a smaller size.

\renewcommand\tabcolsep{0.15cm}
\renewcommand{\arraystretch}{1.50}
\begin{table}[!htbp]
\centering
\caption{The obtained loosely bound-state solutions for the $\Xi_b B$ state with $I(J^P)=0(1/2^-)$. The probabilities $P$ of the various components are given in percent. Here, we note that the channels with probabilities below $5\times 10^{-3}$\% are denoted simply as $\sim0$.}\label{XibB1}
\begin{tabular}{cccc}
\toprule[1pt]
\toprule[1pt]
\multicolumn{4}{c}{Single channel case} \\
\cline{1-4}
$\Lambda$ (GeV) & $E$ (MeV) & $r_{\rm RMS}$ (fm) & $P({}^2S_{1/2})$ \\
0.35&$-2.08$&2.00&100\\
0.45&$-5.86$&1.31&100\\
0.55&$-11.84$&0.99&100 \\
\cline{1-4}
\multicolumn{4}{c}{Coupled channel case}\\
\cline{1-4}
$\Lambda$ (GeV) & $E$ (MeV) & $r_{\rm RMS}$ (fm) & $P(\Xi_b B/\Xi_b B^*/\Xi_b^{\prime} B/\Xi_b^{\prime} B^*/\Xi_b^* B^*)$ \\
0.35&$-2.12$&1.98& 99.98/$\sim0$/$\sim0$/0.01/0.01\\
0.45&$-6.03$&1.30& 99.91/$\sim0$/$\sim0$/0.03/0.06\\
0.55&$-12.22$&0.98&99.77/$\sim0$/$\sim0$/0.09/0.14\\
\bottomrule[1pt]\bottomrule[1pt]
\end{tabular}
\end{table}

When the coupled-channel effects are incorporated, the Hilbert space is extended to include the $\Xi_b B^*$, $\Xi_b^\prime B$, $\Xi_b^\prime B^*$, and $\Xi_b^* B^*$ channels, which are connected to the leading channel $\Xi_b B$ through the off-diagonal transitions. The lower panel of Table \ref{XibB1} shows that the inclusion of these channels only slightly deepens the binding energy, by $0.04$-$0.38$ MeV relative to the single-channel results for a fixed cutoff value. The dominant component remains the $\Xi_b B$ configuration, with probabilities exceeding $99.7\%$ across the entire cutoff range. The admixtures from the other channels are extremely small: the $\Xi_b B^*$ and $\Xi_b^\prime B$ probabilities are practically zero, while the $\Xi_b^\prime B^*$ and $\Xi_b^* B^*$ components together contribute less than $0.3\%$.

\paragraph{$\Xi_b B^*$ system}

We now turn to the bound-state properties for the $\Xi_b B^*$ system, which exhibits a richer dynamical structure than the $\Xi_b B$ system owing to the vector nature of the $B^*$ meson. This system permits two distinct spin-parity assignments, namely $J^P = 1/2^-$ and $3/2^-$, and the corresponding numerical results are compiled in Table \ref{XibB2}.

\begin{table}[!htbp]
\renewcommand{\tabcolsep}{0.10cm}
\renewcommand{\arraystretch}{1.50}
\centering
\caption{The obtained loosely bound-state solutions for the $\Xi_b B^*$ system. The probabilities $P$ of the various components are given in percent.}\label{XibB2}
\begin{tabular}{ccccc}
\toprule[1pt]
\toprule[1pt]
\multirow{10}{*}{$0\left(\frac{1}{2}^-\right)$}&\multicolumn{4}{c}{Single channel case} \\
\cline{2-5}
&$\Lambda$ (GeV) & $E$ (MeV) & $r_{\rm RMS}$ (fm) & $P({}^2S_{1/2}/{}^4D_{1/2})$ \\
&0.35&$-2.10$&1.99&100/0\\
&0.45&$-5.91$&1.31&100/0\\
&0.55&$-11.92$&0.99&100/0\\
\cline{2-5}
&\multicolumn{4}{c}{Coupled channel case}\\
\cline{2-5}
&$\Lambda$ (GeV) & $E$ (MeV) & $r_{\rm RMS}$ (fm) & $P(\Xi_b B^*/\Xi_b^\prime B/\Xi_b^\prime B^*/\Xi_b^* B^*)$ \\
&0.34&$-2.09$&1.99&99.82/0.04/0.10/0.04\\
&0.41&$-5.19$&1.37&99.19/0.12/0.49/0.20\\
&0.48&$-11.17$&0.99&97.05/0.28/1.94/0.73\\
\cline{1-5}
\multirow{10}{*}{$0\left(\frac{3}{2}^-\right)$}&\multicolumn{4}{c}{Single channel case} \\
\cline{2-5}
&$\Lambda$ (GeV) & $E$ (MeV) & $r_{\rm RMS}$ (fm) & $P({}^4S_{3/2}/{}^2D_{3/2}/{}^4D_{3/2})$ \\
&0.35&$-2.10$&1.99&100/0/0\\
&0.45&$-5.91$&1.31&100/0/0\\
&0.55&$-11.92$&0.99&100/0/0\\
\cline{2-5}
&\multicolumn{4}{c}{Coupled channel case}\\
\cline{2-5}
&$\Lambda$ (GeV) & $E$ (MeV) & $r_{\rm RMS}$ (fm) & $P(\Xi_b B^*/\Xi_b^* B/\Xi_b^\prime B^*/\Xi_b^* B^*)$ \\
&0.34&$-2.08$&2.00&99.84/0.03/0.03/0.10\\
&0.41&$-5.13$&1.38&99.31/0.09/0.11/0.49\\
&0.48&$-10.92$&1.00&97.51/0.18/0.42/1.89\\
\bottomrule[1pt]\bottomrule[1pt]
\end{tabular}
\end{table}

For the $\Xi_b B^*$ state with $I(J^P)=0(1/2^-)$, the single-channel calculation restricted to the $\Xi_b B^*({}^2S_{1/2})$ and $\Xi_b B^*({}^4D_{1/2})$ channels yields the binding energies and the RMS radii that are nearly degenerate with those of the $\Xi_b B$ state with $I(J^P)=0(1/2^-)$. This near equality is primarily attributable to the identical interactions in both systems, with the slight difference originating from the marginally larger reduced mass of the $\Xi_b B^*$ system \cite{Li:2014gra,Karliner:2015ina}. Furthermore, the absence of the tensor force in the single-channel treatment naturally suppresses any $D$-wave admixture, leaving the wave function entirely in the $S$-wave sector. Once the coupled-channel effects are incorporated, the Hilbert space expands to include the $\Xi_b^\prime B$, $\Xi_b^\prime B^*$, and $\Xi_b^* B^*$ configurations. At a relatively small cutoff $\Lambda = 0.34$ GeV, the wave function remains overwhelmingly dominated by the $\Xi_b B^*$ component ($99.82\%$), indicating that the coupled channels are only weakly coupled at this cutoff value. As $\Lambda$ increases to $0.48$ GeV, the $\Xi_b B^*$ fraction decreases to $97.05\%$, while the $\Xi_b^\prime B^*$ and $\Xi_b^* B^*$ admixtures grow to $1.94\%$ and $0.73\%$, respectively. This increasing admixture reflects the enhanced transition strengths driven by the stronger effective interaction at larger cutoff values, which promotes mixing among the different channels.

Turning to the $\Xi_b B^*$ state with $I(J^P)=0(3/2^-)$, the single-channel approximation yields results that are fully degenerate with those of the $I(J^P)=0(1/2^-)$ $\Xi_b B^*$ state. This degeneracy is a natural consequence of the absence of the spin-spin interactions in the single-channel case for the $\Xi_b B^*$ system, where the spin multiplicity does not alter the binding properties. Upon switching on the coupled-channel dynamics, the degeneracy is slightly lifted. When the cutoff parameter is set to the same value, the binding energy and the RMS radius differ slightly. Taking $\Lambda = 0.48$ GeV as a reference, the $I(J^P)=0(3/2^-)$ $\Xi_b B^*$ coupled system has a slightly shallower binding energy ($E = -10.92$ MeV) and a marginally larger RMS radius ($r_{\rm RMS} = 1.00$ fm) compared to its $I(J^P)=0(1/2^-)$ counterpart ($E = -11.17$ MeV and $r_{\rm RMS} = 0.99$ fm). It is worth noting that in the $I(J^P)=0(3/2^-)$ $\Xi_b B^*$ coupled system, the $\Xi_b^* B^*$ admixture reaches $1.89\%$ at $\Lambda = 0.48$ GeV, significantly larger than the $0.73\%$ observed in the $I(J^P)=0(1/2^-)$ $\Xi_b B^*$ coupled system. This slight difference, though modest in magnitude, signals the emergence of the spin-dependent forces arising from the coupled-channel mixings. Overall, the $\Xi_b B^*$ system demonstrates that the coupled-channel effects, while quantitatively moderate, introduce nontrivial spin-dependent structures that distinguish the $I(J^P)=0(1/2^-)$ and $0(3/2^-)$ $\Xi_b B^*$ bound states.

\paragraph{$\Xi_b^\prime B$ system}

We next examine the bound-state properties for the $\Xi_b^\prime B$ state with quantum numbers $I(J^P)=0(1/2^-)$. The results are displayed in Table \ref{XibB3}. In the single-channel calculation, the $\Xi_b^\prime B$ state with $I(J^P)=0(1/2^-)$ forms the loosely bound states for the cutoff parameters $\Lambda = 0.36$, $0.46$, and $0.56$ GeV, with the binding energies $E = -2.01$, $-5.59$, and $-11.20$ MeV, respectively.

\renewcommand\tabcolsep{0.32cm}
\renewcommand{\arraystretch}{1.50}
\begin{table}[!htbp]
\centering
\caption{The obtained loosely bound-state solutions for the $\Xi_b^{\prime} B$ state with $I(J^P)=0(1/2^-)$. The probabilities $P$ of the various components are given in percent.}\label{XibB3}
\begin{tabular}{cccc}
\toprule[1pt]
\toprule[1pt]
\multicolumn{4}{c}{Single channel case} \\
\cline{1-4}
$\Lambda$ (GeV) & $E$ (MeV) & $r_{\rm RMS}$ (fm) & $P({}^2S_{1/2})$ \\
0.36&$-2.01$&2.00&100\\
0.46&$-5.59$&1.32&100\\
0.56&$-11.20$&1.00&100 \\
\cline{1-4}
\multicolumn{4}{c}{Coupled channel case}\\
\cline{1-4}
$\Lambda$ (GeV) & $E$ (MeV) & $r_{\rm RMS}$ (fm) & $P(\Xi_b^{\prime} B/\Xi_b^{\prime} B^*/\Xi_b^* B^*)$ \\
0.36&$-2.16$&1.94&99.68/0.26/0.06\\
0.44&$-5.29$&1.35&98.52/1.29/0.19\\
0.52&$-10.72$&0.99&89.64/10.28/0.08\\
\bottomrule[1pt]\bottomrule[1pt]
\end{tabular}
\end{table}

When the coupled-channel dynamics are included, the $\Xi_b^\prime B^*$ and $\Xi_b^* B^*$ channels are incorporated. The effect of channel coupling is considerably more pronounced than in the $\Xi_b B$ coupled system with $I(J^P)=0(1/2^-)$. At $\Lambda = 0.36$ GeV, the binding energy becomes $-2.16$ MeV, slightly enhanced compared to the single-channel value, and the wave function remains predominantly $\Xi_b^\prime B$ with $99.68\%$ probability. As $\Lambda$ increases to $0.52$ GeV, the $\Xi_b^\prime B$ component decreases to $89.64\%$, while the $\Xi_b^\prime B^*$ admixture grows substantially to $10.28\%$. This significant mixing is in sharp contrast to the $\Xi_b B$ coupled system with $I(J^P)=0(1/2^-)$, where all coupled-channel admixtures remained well below $1\%$. The enhanced coupling between the $\Xi_b^\prime B$ and $\Xi_b^\prime B^*$ channels can be traced back to the spin structure of the $\Xi_b^\prime$ baryon. Unlike the $\Xi_b$, which has light-quark spin $S_{\ell} = 0$, the $\Xi_b^\prime$ belongs to the sextet representation and carries $S_{\ell}=1$. This additional light-spin degree of freedom gives rise to the spin-spin and tensor interaction terms that are absent or suppressed in the anti-triplet case, thereby opening stronger off-diagonal transitions between the $\Xi_b^\prime B$ and $\Xi_b^\prime B^*$ channels. Nevertheless, even at the largest cutoff considered, the $\Xi_b^\prime B$ component still dominates the wave function with nearly $90\%$ probability, and the $\Xi_b^* B^*$ admixture remains negligible ($0.08\%$). Its relatively shallow binding and significant dominance make it another promising candidate of the hidden-bottom molecular pentaquark with strangeness.

\paragraph{$\Xi_b^* B$ system}

We now proceed to analyze the bound-state properties for the $\Xi_b^* B$ state with quantum numbers $I(J^P) = 0(3/2^-)$. The corresponding results are listed in Table \ref{XibB4}. In the single-channel calculation, restricted to the $\Xi_b^* B({}^4S_{3/2})$ and $\Xi_b^* B({}^4D_{3/2})$ channels, the binding energies are found to be $E = -2.02$, $-5.60$, and $-11.23$ MeV for the cutoff parameters $\Lambda = 0.36$, $0.46$, and $0.56$ GeV, respectively. The ${}^4D_{3/2}$ partial wave does not contribute because the tensor force, which would otherwise induce the $S$-$D$ wave mixing effects, is absent in the $\Xi_b^* B$ interaction.

\renewcommand\tabcolsep{0.32cm}
\renewcommand{\arraystretch}{1.50}
\begin{table}[!htbp]
\centering
\caption{The obtained loosely bound-state solutions for the $\Xi_b^* B$ state with $I(J^P)=0(3/2^-)$. The probabilities $P$ of the various components are given in percent.}\label{XibB4}
\begin{tabular}{cccc}
\toprule[1pt]
\toprule[1pt]
\multicolumn{4}{c}{Single channel case} \\
\cline{1-4}
$\Lambda$ (GeV) & $E$ (MeV) & $r_{\rm RMS}$ (fm) & $P({}^4S_{3/2}/{}^4D_{3/2})$ \\
0.36&$-2.02$&1.99&100/0\\
0.46&$-5.60$&1.32&100/0\\
0.56&$-11.23$&1.00&100/0\\
\cline{1-4}
\multicolumn{4}{c}{Coupled channel case}\\
\cline{1-4}
$\Lambda$ (GeV) & $E$ (MeV) & $r_{\rm RMS}$ (fm) & $P(\Xi_b^* B/\Xi_b^{\prime} B^*/\Xi_b^* B^*)$ \\
0.36&$-2.17$&1.94&99.67/0.07/0.26\\
0.46&$-6.26$&1.26&98.49/0.09/1.42\\
0.56&$-13.42$&0.91&91.85/0.22/7.93\\
\bottomrule[1pt]\bottomrule[1pt]
\end{tabular}
\end{table}

When the coupled-channel effects are included, the $\Xi_b^\prime B^*$ and $\Xi_b^* B^*$ channels are incorporated into the calculation. At $\Lambda = 0.36$ GeV, the binding energy is slightly enhanced to $-2.17$ MeV compared to the single-channel value, and the wave function remains predominantly $\Xi_b^* B$ with $99.67\%$ probability. However, as $\Lambda$ increases to $0.56$ GeV, the $\Xi_b^* B$ component decreases to $91.85\%$, while the $\Xi_b^* B^*$ admixture grows substantially to $7.93\%$. The $\Xi_b^\prime B^*$ component remains small throughout, at most $0.22\%$. This substantial coupled-channel mixing leads to a deepening of the binding energy by $2.19$ MeV at $\Lambda = 0.56$ GeV compared to the single-channel result.

\paragraph{$\Xi_b^\prime B^*$ system}

We next analyze the bound-state properties for the $\Xi_b^\prime B^*$ system, which supports two distinct spin-parity assignments, namely $J^P = 1/2^-$ and $3/2^-$. The corresponding numerical results are compiled in Table \ref{XibB5}.

\renewcommand\tabcolsep{0.16cm}
\renewcommand{\arraystretch}{1.50}
\begin{table}[!htbp]
\centering
\caption{The obtained loosely bound-state solutions for the $\Xi_b^{\prime} B^*$ system. The probabilities $P$ of the various components are given in percent.}\label{XibB5}
\begin{tabular}{ccccc}
\toprule[1pt]
\toprule[1pt]
\multirow{10}{*}{$0\left(\frac{1}{2}^-\right)$}&\multicolumn{4}{c}{Single channel case} \\
\cline{2-5}
&$\Lambda$ (GeV) & $E$ (MeV) & $r_{\rm RMS}$ (fm) & $P({}^2S_{1/2}/{}^4D_{1/2})$ \\
&0.28&$-1.97$&2.04&99.91/0.09\\
&0.34&$-5.48$&1.36&99.90/0.10\\
&0.40&$-11.76$&0.99&99.91/0.09\\
\cline{2-5}
&\multicolumn{4}{c}{Coupled channel case}\\
\cline{2-5}
&$\Lambda$ (GeV) & $E$ (MeV) & $r_{\rm RMS}$ (fm) & $P(\Xi_b^{\prime} B^*/\Xi_b^* B^*)$ \\
&0.28&$-2.01$&2.02&99.77/0.23\\
&0.34&$-5.77$& 1.32& 98.52/1.48\\
&0.39&$-11.52$&0.99&94.65/5.35\\
\cline{1-5}
\multirow{10}{*}{$0\left(\frac{3}{2}^-\right)$}&\multicolumn{4}{c}{Single channel case} \\
\cline{2-5}
&$\Lambda$ (GeV) & $E$ (MeV) & $r_{\rm RMS}$ (fm) & $P({}^4S_{3/2}/{}^2D_{3/2}/{}^4D_{3/2})$ \\
&0.44&$-2.02$&2.03&98.72/0.24/1.04\\
&0.52&$-3.50$&1.65&98.23/0.34/1.43\\
&0.60&$-5.21$&1.42&97.70/0.45/1.85\\
\cline{2-5}
&\multicolumn{4}{c}{Coupled channel case}\\
\cline{2-5}
&$\Lambda$ (GeV) & $E$ (MeV) & $r_{\rm RMS}$ (fm) & $P(\Xi_b^{\prime} B^*/\Xi_b^* B^*)$ \\
&0.40&$-2.03$&1.94&94.59/5.41\\
&0.45&$-4.58$&1.31&79.17/20.83\\
&0.49&$-8.82$&0.95&58.08/41.92\\
\bottomrule[1pt]\bottomrule[1pt]
\end{tabular}
\end{table}

For the $\Xi_b^\prime B^*$ state with $I(J^P)=0(1/2^-)$, the single-channel calculation restricted to the $\Xi_b^\prime B^*({}^2S_{1/2})$ and $\Xi_b^\prime B^*({}^4D_{1/2})$ channels yields the binding energies of $E = -1.97$, $-5.48$, and $-11.76$ MeV for the cutoff parameters $\Lambda = 0.28$, $0.34$, and $0.40$ GeV, respectively. The wave function remains overwhelmingly $S$-wave, with only a tiny ${}^4D_{1/2}$ admixture of about $0.1\%$, indicating that the tensor force is suppressed in this channel. When the $\Xi_b^* B^*$ channel is incorporated into the coupled-channel framework, the mixing effects are found to be moderate. At the largest cutoff considered, the $\Xi_b^\prime B^*$ component remains dominant at $94.65\%$, while the $\Xi_b^* B^*$ admixture reaches $5.35\%$.

For the $\Xi_b^\prime B^*$ state with $I(J^P)=0(3/2^-)$, a markedly different behavior emerges. In the single-channel calculation, larger cutoff values are needed to generate the loosely bound-state solutions: $\Lambda = 0.44$, $0.52$, and $0.60$ GeV yield the binding energies of $-2.02$, $-3.50$, and $-5.21$ MeV, respectively. More importantly, the $D$-wave admixtures are no longer negligible in this channel. The combined contribution of the ${}^2D_{3/2}$ and ${}^4D_{3/2}$ partial waves grows from about $1.3\%$ at $\Lambda = 0.44$ GeV to $2.3\%$ at $\Lambda = 0.60$ GeV, reflecting the increasingly prominent role of the tensor force as the cutoff value expands. This contrasts with the $\Xi_b^\prime B^*$ state with $I(J^P) =0(1/2^-)$, where the tensor-induced $D$-wave components remain essentially invisible throughout the entire cutoff range. When the coupled-channel dynamics with the $\Xi_b^* B^*$ channel are switched on, the $\Xi_b^\prime B^*$ coupled system with $I(J^P)=0(3/2^-)$ exhibits particularly strong mixing. At $\Lambda = 0.40$ GeV, the binding energy is $-2.03$ MeV, with the $\Xi_b^\prime B^*$ component accounting for $94.59\%$ and the $\Xi_b^* B^*$ admixture at $5.41\%$. As $\Lambda$ increases to $0.49$ GeV, the $\Xi_b^\prime B^*$ and $\Xi_b^* B^*$ probabilities become nearly equal, at $58.08\%$ and $41.92\%$, respectively. The binding energy at this cutoff is $-8.82$ MeV. This behavior is in stark contrast to the $\Xi_b^\prime B^*$ coupled system with $I(J^P)=0(1/2^-)$, where the $\Xi_b^* B^*$ admixture reaches only $5.35\%$ at the largest cutoff.

\paragraph{$\Xi_b^* B^*$ system}

In the following, we discuss the bound-state properties for the $\Xi_b^* B^*$ system, which involves one spin-$3/2$ baryon and one spin-$1$ meson. This configuration supports three distinct spin-parity channels: $J^P = 1/2^-$, $3/2^-$, and $5/2^-$. The corresponding numerical results are presented in Table \ref{XibB6}.

\renewcommand\tabcolsep{0.08cm}
\renewcommand{\arraystretch}{1.50}
\begin{table}[!htbp]
\centering
\caption{The obtained loosely bound-state solutions for the $\Xi_b^* B^*$ system. The probabilities $P$ of the various components are given in percent.}\label{XibB6}
\begin{tabular}{ccccc}
\toprule[1pt]
\toprule[1pt]
\multirow{4}{*}{$0\left(\frac{1}{2}^-\right)$}&$\Lambda$ (GeV) & $E$ (MeV) & $r_{\rm RMS}$ (fm) & $P({}^2S_{1/2}/{}^4D_{1/2}/{}^6D_{1/2})$ \\
&0.27 & $-2.05$ & 2.02 & 99.83 / 0.10 / 0.07\\
&0.33&$-6.00$ & 1.32 & 99.83 / 0.11 / 0.06\\
&0.38 & $-11.80$ & 1.00 & 99.84 / 0.10 / 0.06\\
\cline{1-5}
\multirow{4}{*}{$0\left(\frac{3}{2}^-\right)$}&$\Lambda$ (GeV) & $E$ (MeV) & $r_{\rm RMS}$ (fm) & $P({}^4S_{3/2}/{}^2D_{3/2}/{}^4D_{3/2}/{}^6D_{3/2})$ \\
&0.31 & $-2.02$ & 2.02 & 99.69 / 0.09 / 0.20 / 0.02\\
&0.39&$-6.11$ & 1.29 & 99.62 / 0.11 / 0.24 / 0.03\\
&0.46 & $-12.33$ & 0.97 & 99.60 / 0.12 / 0.25 / 0.03\\
\cline{1-5}
\multirow{4}{*}{$0\left(\frac{5}{2}^-\right)$}&$\Lambda$ (GeV) & $E$ (MeV) & $r_{\rm RMS}$ (fm) & $P({}^6S_{5/2}/{}^2D_{5/2}/{}^4D_{5/2}/{}^6D_{5/2})$ \\
&0.51 & $-2.05$ & 2.06 & 97.40 / 0.19 / 0.09 / 2.32\\
&0.56&$-2.69$ & 1.86 & 96.81 / 0.23 / 0.11 / 2.85\\
&0.60 & $-3.24$ & 1.74 & 96.32 / 0.26 / 0.13 / 3.29\\
\bottomrule[1pt]\bottomrule[1pt]
\end{tabular}
\end{table}

For the $\Xi_b^* B^*$ state with $I(J^P)=0(1/2^-)$, the single-channel calculation yields the binding energies of $E = -2.05$, $-6.00$, and $-11.80$ MeV for the cutoff parameters $\Lambda = 0.27$, $0.33$, and $0.38$ GeV, respectively. The wave function is overwhelmingly dominated by the ${}^2S_{1/2}$ component, with the ${}^4D_{1/2}$ and ${}^6D_{1/2}$ admixtures remaining at the level of about $0.1\%$ or less. This pattern closely resembles that observed in the $\Xi_b^\prime B^*$ state for the same spin-parity, suggesting that the tensor force is similarly suppressed in this channel.

The $\Xi_b^* B^*$ state with $I(J^P)=0(3/2^-)$ yields the binding energies of $-2.02$, $-6.11$, and $-12.33$ MeV for $\Lambda = 0.31$, $0.39$, and $0.46$ GeV, respectively. The dominant configuration is the ${}^4S_{3/2}$ partial wave, with small admixtures from the ${}^2D_{3/2}$, ${}^4D_{3/2}$, and ${}^6D_{3/2}$ partial waves. The total $D$-wave probability remains below $0.4\%$ across the entire cutoff range, indicating that the tensor force effects, while slightly more noticeable than in the $I(J^P)=0(1/2^-)$ $\Xi_b^* B^*$ bound state, still play a relatively minor role in this bound state.

The $\Xi_b^* B^*$ state with $I(J^P)=0(5/2^-)$ exhibits distinct features that set it apart from the lower-spin cases. For the $\Xi_b^* B^*$ state with $I(J^P)=0(5/2^-)$, the binding energies are $-2.05$, $-2.69$, and $-3.24$ MeV for $\Lambda = 0.51$, $0.56$, and $0.60$ GeV, respectively. Notably, this bound state requires substantially larger cutoff parameters to achieve binding energies comparable to those of the lower-spin states. This behavior can be understood from the interaction potentials: the spin couplings enter with different combinations of the interaction strengths, resulting in a weaker effective attraction that must be compensated by a larger cutoff value for the $\Xi_b^* B^*$ state with $I(J^P)=0(5/2^-)$. Moreover, the $D$-wave admixtures in this bound state are considerably more pronounced. At larger cutoff values, the ${}^6D_{5/2}$ component reaches $3.29\%$, and the total $D$-wave probability exceeds $3.6\%$, which is an order of magnitude larger than in the $\Xi_b^* B^*$ bound states with $I(J^P)=0(1/2^-)$ and $0(3/2^-)$.

\begin{figure}[htbp]
\includegraphics[scale=0.60]{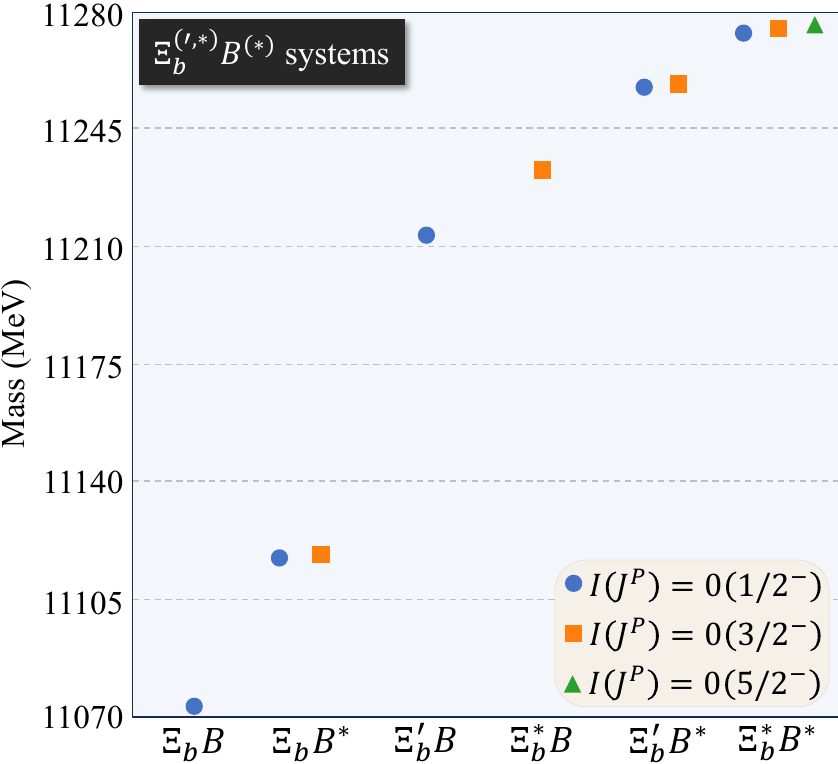}
\caption{Predicted mass spectra for the promising candidates of the $\Xi_b^{(\prime,\,*)} B^{(*)}$-type singly-strange hidden-bottom molecular pentaquarks.}\label{XibBmassspectra}
\end{figure}

Based on our systematic calculations, we identify a set of promising candidates for the $\Xi_b^{(\prime,\,*)} B^{(*)}$-type singly-strange hidden-bottom molecular pentaquarks, i.e.,
\begin{itemize}
\item $\Xi_b B$ state with $I(J^P) = 0(1/2^-)$;
\item $\Xi_b B^*$ states with $I(J^P) = 0(1/2^-)$ and $0(3/2^-)$;
\item $\Xi_b^\prime B$ state with $I(J^P) = 0(1/2^-)$;
\item $\Xi_b^* B$ state with $I(J^P) = 0(3/2^-)$;
\item $\Xi_b^\prime B^*$ states with $I(J^P) = 0(1/2^-)$ and $0(3/2^-)$;
\item $\Xi_b^* B^*$ states with $I(J^P) = 0(1/2^-)$, $0(3/2^-)$, and $0(5/2^-)$.
\end{itemize}
The predicted mass spectra for these candidates are displayed in Fig. \ref{XibBmassspectra}.

\subsection{Singly-strange double-bottom $\Xi_b^{(\prime,\,*)}\bar B^{(*)}$ systems}

In this subsection, we extend our study to discuss the bound-state properties for the singly-strange double-bottom pentaquark systems composed of $\Xi_b^{(\prime,\,*)}$ baryons and $\bar B^{(*)}$ mesons. As detailed in Sec. \ref{theory}, the effective potentials for these systems can be obtained from those of the corresponding  singly-strange hidden-bottom $\Xi_b^{(\prime,\,*)}B^{(*)}$ partners by applying the $G$-parity rule \cite{Klempt:2002ap}. This symmetry operation plays an essential role in shaping the effective interactions: it reverses the sign of the contributions from the mesons with negative $G$-parity, most importantly the $\pi$ and $\omega$, while leaving the contributions from positive $G$-parity mesons, such as the $\sigma$, $\eta$, and $\rho$, unchanged. As a consequence, the relative signs of the $\pi$ and $\omega$ exchange terms are rearranged, which can modify the attraction or repulsion in each $I(J^P)$ channel as compared with the singly-strange hidden-bottom $\Xi_b^{(\prime,\,*)}B^{(*)}$ systems. In the following, we present the bound-state properties for the singly-strange double-bottom  $\Xi_b^{(\prime,\,*)}\bar B^{(*)}$ systems.

\paragraph{$\Xi_b \bar B$ and $\Xi_b \bar B^*$ systems}

We first present the bound-state properties for the double-bottom $\Xi_b \bar B$ and $\Xi_b \bar B^*$ systems, which are summarized in Table \ref{XibbarB1}. Compared with their hidden-bottom counterparts, these systems exhibit a notable feature: the binding occurs at the considerably smaller cutoff parameters.

\renewcommand\tabcolsep{0.05cm}
\renewcommand{\arraystretch}{1.50}
\begin{table}[!htbp]
\centering
\caption{The obtained loosely bound-state solutions for the $\Xi_b \bar B$ and $\Xi_b \bar B^*$ systems. The probabilities $P$ of the various components are given in percent. Here, we note that the channels with probabilities below $5\times 10^{-3}$\% are denoted simply as $\sim0$.}\label{XibbarB1}
\begin{tabular}{ccccc}
\toprule[1pt]\toprule[1pt]
States &$\Lambda$ (GeV) & $E$ (MeV) & $r_{\rm RMS}$ (fm) & $P$ (Components) \\
\cline{1-5}
\multirow{8}{*}{$\Xi_b \bar B[0(\frac{1}{2}^-)]$} & \multicolumn{4}{c}{Single channel: $P({}^2S_{1/2})$} \\
&0.25&$-2.14$&2.17& 100\\
&0.33&$-6.57$&1.38& 100\\
&0.41&$-14.09$&1.01& 100\\
\cline{2-5}
& \multicolumn{4}{c}{Coupled channel: $P(\Xi_b \bar B/\Xi_b \bar B^*/\Xi_b' \bar B/\Xi_b' \bar B^*/\Xi_b^* \bar B^*)$} \\
&0.25&$-2.23$&2.13&99.96/$\sim0$/$\sim0$/0.01/0.03\\
&0.32&$-6.52$&1.38&99.69/$\sim0$/$\sim0$/0.11/0.20\\
&0.38&$-13.65$&1.01&98.65/$\sim0$/$\sim0$/0.50/0.85\\
\cline{1-5}
\multirow{8}{*}{$\Xi_b \bar B^*[0(\frac{1}{2}^-)]$} & \multicolumn{4}{c}{Single channel: $P({}^2S_{1/2}/{}^4D_{1/2})$} \\
&0.25&$-2.16$&2.16&100/0\\
&0.33&$-6.61$&1.38&100/0\\
&0.41&$-14.16$&1.01&100/0\\
\cline{2-5}
&\multicolumn{4}{c}{Coupled channel: $P(\Xi_b \bar B^*/\Xi_b' \bar B/\Xi_b' \bar B^*/\Xi_b^* \bar B^*)$} \\
&0.25&$-2.24$&2.13&99.93/0.06/0.01/$\sim0$\\
&0.32&$-6.39$&1.39&99.55/0.44/0.01/$\sim0$\\
&0.39&$-14.27$&0.99&97.84/2.11/0.03/0.02\\
\cline{1-5}
\multirow{8}{*}{$\Xi_b \bar B^*[0(\frac{3}{2}^-)]$} & \multicolumn{4}{c}{Single channel: $P({}^4S_{3/2}/{}^2D_{3/2}/{}^4D_{3/2})$} \\
&0.25&$-2.16$&2.16&100/0/0\\
&0.33&$-6.61$&1.38&100/0/0\\
&0.41&$-14.16$&1.01&100/0/0\\
\cline{2-5}
& \multicolumn{4}{c}{Coupled channel: $P(\Xi_b \bar B^*/\Xi_b^* \bar B/\Xi_b' \bar B^*/\Xi_b^* \bar B^*)$} \\
&0.25&$-2.24$&2.13&99.95/0.05/$\sim0$/$\sim0$\\
&0.32&$-6.35$&1.39&99.64/0.34/0.01/0.01\\
&0.39&$-14.04$&1.00&98.35/1.59/0.03/0.03\\
\bottomrule[1pt]\bottomrule[1pt]
\end{tabular}
\end{table}

For the $\Xi_b \bar B$ state with $I(J^P)=0(1/2^-)$, the single-channel calculation yields the binding energies of $-2.14$, $-6.57$, and $-14.09$ MeV for $\Lambda = 0.25$, $0.33$, and $0.41$ GeV, respectively. These values should be contrasted with those of the $\Xi_b B$ state with $I(J^P)=0(1/2^-)$, where a cutoff around $0.33$ GeV only produces a binding of about $1.58$ MeV. This is due to the $G$-parity transformation flips the sign of the $\omega$-exchange contribution, which is repulsive in the hidden-bottom case and would be expected to reduce the attraction. When the coupled-channel effects are switched on, the $\Xi_b \bar B$ component continues to dominate the wave function, with probabilities of $99.96\%$, $99.69\%$, and $98.65\%$ at $\Lambda = 0.25$, $0.32$, and $0.38$ GeV, respectively. The admixtures from the $\Xi_b^\prime \bar B^*$ and $\Xi_b^* \bar B^*$ channels remain small, reaching at most $0.50\%$ and $0.85\%$, while the $\Xi_b \bar B^*$ and $\Xi_b^\prime \bar B$ channels are essentially negligible. This pattern closely resembles that of the hidden-bottom $\Xi_b B$ coupled system with $I(J^P)=0(1/2^-)$.

The $\Xi_b \bar B^*$ system admits both $J^P = 1/2^-$ and $3/2^-$ configurations. In the single-channel approximation, the $I(J^P)=0(1/2^-)$ and $0(3/2^-)$ $\Xi_b \bar B^*$ bound states are degenerate, yielding identical binding energies of $-2.16$, $-6.61$, and $-14.16$ MeV for $\Lambda = 0.25$, $0.33$, and $0.41$ GeV, respectively. These values are very close to those of the $\Xi_b \bar B$ state with $I(J^P)=0(1/2^-)$. For the $I(J^P)=0(1/2^-)$ $\Xi_b \bar B^*$ coupled system, the $\Xi_b \bar B^*$ component remains dominant but shows a more pronounced decrease with increasing cutoff compared to the $\Xi_b \bar B$ coupled system with $I(J^P)=0(1/2^-)$. At $\Lambda = 0.25$ GeV, the probability is $99.93\%$. At $\Lambda = 0.39$ GeV, it drops to $97.84\%$. Meanwhile, the $\Xi_b^\prime \bar B$ channel grows to $2.11\%$, and the $\Xi_b^\prime \bar B^*$ and $\Xi_b^* \bar B^*$ channels appear at the level of $0.03\%$ or less. For the $I(J^P)=0(3/2^-)$ $\Xi_b \bar B^*$ coupled system, the coupled-channel effects are somewhat milder. The $\Xi_b \bar B^*$ component remains high, with probabilities of $99.95\%$, $99.64\%$, and $98.35\%$ at $\Lambda = 0.25$, $0.32$, and $0.39$ GeV, respectively. The $\Xi_b^* \bar B$ admixture reaches $1.59\%$ at the largest cutoff, while the other channels are negligible.

\paragraph{$\Xi_b^\prime \bar B$ and $\Xi_b^* \bar B$ systems}

We now turn to discuss the bound-state properties for the $\Xi_b^\prime \bar B$ and $\Xi_b^* \bar B$ systems, which represent the double-bottom analogues of the hidden-bottom $\Xi_b^\prime B$ and $\Xi_b^* B$ systems studied previously. The bound-state properties for the $\Xi_b^\prime \bar B$ and $\Xi_b^* \bar B$ systems are collected in Table \ref{XibbarB2}.

\renewcommand\tabcolsep{0.16cm}
\renewcommand{\arraystretch}{1.50}
\begin{table}[!htbp]
\centering
\caption{The obtained loosely bound-state solutions for the $\Xi_b^{\prime} \bar B$ and $\Xi_b^* \bar B$ systems. The probabilities $P$ of the various components are given in percent.}\label{XibbarB2}
\begin{tabular}{cccccc}
\toprule[1pt]
\toprule[1pt]
States &$\Lambda$ (GeV) & $E$ (MeV) & $r_{\rm RMS}$ (fm) & $P$ (Components) \\
\cline{1-5}
\multirow{8}{*}{$\Xi_b^{\prime} \bar B[0(\frac{1}{2}^-)]$} & \multicolumn{4}{c}{Single channel: $P({}^2S_{1/2})$} \\
&0.25&$-1.99$&2.21&100\\
&0.34&$-6.89$&1.34&100\\
&0.42&$-14.43$&0.99&100\\
\cline{2-5}
& \multicolumn{4}{c}{Coupled channel: $P(\Xi_b' \bar B/\Xi_b' \bar B^*/\Xi_b^* \bar B^*)$} \\
&0.25&$-2.11$&2.16&99.79/0.17/0.04\\
&0.31&$-5.63$&1.44&98.69/1.03/0.28\\
&0.37&$-12.68$&1.01&94.53/4.28/1.19\\
\cline{1-5}
\multirow{8}{*}{$\Xi_b^* \bar B[0(\frac{3}{2}^-)]$}& \multicolumn{4}{c}{Single channel: $P({}^4S_{3/2}/{}^4D_{3/2})$} \\
&0.25&$-1.99$&2.21&100/0\\
&0.34&$-6.91$&1.34&100/0\\
&0.42&$-14.46$&0.99&100/0\\
\cline{2-5}
&\multicolumn{4}{c}{Coupled channel: $P(\Xi_b^* \bar B/\Xi_b' \bar B^*/\Xi_b^* \bar B^*)$} \\
&0.25&$-2.14$&2.14&99.67/0.12/0.21\\
&0.31&$-5.80$&1.42&97.97/0.71/1.32\\
&0.37&$-13.25$&0.99&92.32/2.46/5.22\\
\bottomrule[1pt]\bottomrule[1pt]
\end{tabular}
\end{table}

For the $\Xi_b^\prime \bar B$ state with $I(J^P)=0(1/2^-)$, the single-channel calculation yields the binding energies of $-1.99$, $-6.89$, and $-14.43$ MeV for the cutoff parameters $\Lambda = 0.25$, $0.34$, and $0.42$ GeV, respectively. These values are appreciably deeper than those of the corresponding hidden-bottom $\Xi_b^\prime B$ state with $I(J^P)=0(1/2^-)$ at comparable cutoff values. To illustrate, at $\Lambda \approx 0.36$ GeV, the $\Xi_b^\prime B$ state with $I(J^P)=0(1/2^-)$ binds by only about $2$ MeV, while the $\Xi_b^\prime \bar B$ state with $I(J^P)=0(1/2^-)$ already reaches a binding of roughly $8$ MeV. This enhancement again originates from the $G$-parity rule \cite{Klempt:2002ap}, which converts the $\omega$-exchange contribution from repulsive to attractive without altering the $\sigma$- and $\rho$-exchange terms. The combined effect is a gain in the attractive strength, leading to more robust bound states at smaller cutoff values. When the coupled-channel effects are switched on by including the $\Xi_b^\prime \bar B^*$ and $\Xi_b^* \bar B^*$ channels, the $\Xi_b^\prime \bar B$ component remains the dominant one, with probabilities of $99.79\%$, $98.69\%$, and $94.53\%$ at $\Lambda = 0.25$, $0.31$, and $0.37$ GeV, respectively. The $\Xi_b^\prime \bar B^*$ admixture grows from $0.17\%$ to $4.28\%$ as the cutoff increases, while the $\Xi_b^* \bar B^*$ fraction reaches $1.19\%$ at the largest cutoff.

For the $\Xi_b^* \bar B$ state with $I(J^P)=0(3/2^-)$, the single-channel calculation gives the binding energies of $-1.99$, $-6.91$, and $-14.46$ MeV for $\Lambda = 0.25$, $0.34$, and $0.42$ GeV, respectively. In the coupled-channel calculation, where the $\Xi_b^\prime \bar B^*$ and $\Xi_b^* \bar B^*$ channel are included, the $\Xi_b^* \bar B$ component remains overwhelmingly dominant, with probabilities of $99.67\%$, $97.97\%$, and $92.32\%$ at $\Lambda = 0.25$, $0.31$, and $0.37$ GeV, respectively. The $\Xi_b^\prime \bar B^*$ admixture reaches $2.46\%$, while the $\Xi_b^* \bar B^*$ admixture reaches $5.22\%$ at the largest cutoff. This behavior mirrors that of the hidden-bottom $\Xi_b^* B$ coupled system with $I(J^P)=0(3/2^-)$, where the dominant channel also retained more than $90\%$ probability.

\paragraph{$\Xi_b^\prime \bar B^*$ and $\Xi_b^* \bar B^*$ systems}

In the following, we discuss the bound-state properties for the $\Xi_b^\prime \bar B^*$ and $\Xi_b^* \bar B^*$ systems, which include the highest-spin configurations in the discussed double-bottom sector. The numerical results are compiled in Table \ref{XibbarB3}.

\renewcommand\tabcolsep{0.06cm}
\renewcommand{\arraystretch}{1.50}
\begin{table}[!htbp]
\centering
\caption{The obtained loosely bound-state solutions for the $\Xi_b' \bar B^*$ and $\Xi_b^* \bar B^*$ systems. The probabilities $P$ of the various components are given in percent.}\label{XibbarB3}
\begin{tabular}{cccccc}
\toprule[1pt]
\toprule[1pt]
Systems&$I(J^P)$ &$\Lambda$ (GeV) & $E$ (MeV) & $r_{\rm RMS}$ (fm) & $P$ (Components) \\
\multirow{16}{*}{$\Xi_b' \bar B^*$} & \multirow{8}{*}{$0(\frac{1}{2}^-)$} & \multicolumn{4}{c}{Single channel: $P({}^2S_{1/2}/{}^4D_{1/2})$} \\
& &0.27&$-2.23$&2.08&99.73/0.27\\
& &0.34&$-6.16$&1.36&99.50/0.50\\
& &0.40&$-12.12$&1.03&99.26/0.74\\
\cline{3-6}
& & \multicolumn{4}{c}{Coupled channel: $P(\Xi_b' \bar B^*/\Xi_b^* \bar B^*)$} \\
& &0.27&$-2.19$&2.09&99.99/0.01\\
& &0.34&$-6.03$&1.37&99.97/0.03\\
& &0.40&$-11.82$&1.03&99.98/0.02\\
\cline{2-6}
& \multirow{8}{*}{$0(\frac{3}{2}^-)$} & \multicolumn{4}{c}{Single channel: $P({}^4S_{3/2}/{}^2D_{3/2}/{}^4D_{3/2})$} \\
& &0.25&$-2.32$&2.11&99.59/0.08/0.33\\
& &0.33&$-6.95$&1.37&99.12/0.18/0.70\\
& &0.42&$-15.69$&1.01&98.26/0.36/1.38\\
\cline{3-6}
& & \multicolumn{4}{c}{Coupled channel: $P(\Xi_b' \bar B^*/\Xi_b^* \bar B^*)$} \\
& &0.25&$-2.27$&2.11&99.97/0.03\\
& &0.33&$-6.70$&1.38&99.93/0.07\\
& &0.42&$-14.79$&1.01&99.94/0.06\\
\cline{1-6}
\multirow{12}{*}{$\Xi_b^* \bar B^*$} & \multirow{4}{*}{$0(\frac{1}{2}^-)$}& \multicolumn{4}{c}{Single channel: $P({}^2S_{1/2}/{}^4D_{1/2}/{}^6D_{1/2})$} \\
& &0.27&$-2.13$&2.12&99.42/0.32/0.26\\
& &0.34&$-6.11$&1.37&98.94/0.56/0.50\\
& &0.40&$-12.42$&1.02&98.48/0.79/0.73\\
\cline{2-6}
& \multirow{4}{*}{$0(\frac{3}{2}^-)$}& \multicolumn{4}{c}{Single channel: $P({}^4S_{3/2}/{}^2D_{3/2}/{}^4D_{3/2}/{}^6D_{3/2})$} \\
& &0.26&$-2.17$&2.13&99.59/0.13/0.25/0.03\\
& &0.34&$-6.75$&1.34&99.18/0.25/0.50/0.07\\
& &0.41&$-13.87$&1.00&98.72/0.40/0.77/0.11\\
\cline{2-6}
& \multirow{4}{*}{$0(\frac{5}{2}^-)$}& \multicolumn{4}{c}{Single channel: $P({}^6S_{5/2}/{}^2D_{5/2}/{}^4D_{5/2}/{}^6D_{5/2})$} \\
& &0.24&$-2.07$&2.21&99.60/0.05/0.01/0.34\\
& &0.33&$-7.24$&1.36&99.05/0.11/0.04/0.80\\
& &0.42&$-15.98$&1.02&98.14/0.23/0.07/1.56\\
\bottomrule[1pt]\bottomrule[1pt]
\end{tabular}
\end{table}

For the $\Xi_b^\prime \bar B^*$ state with $I(J^P)=0(1/2^-)$, the single-channel calculation yields the binding energies of $-2.23$, $-6.16$, and $-12.12$ MeV at $\Lambda = 0.27$, $0.34$, and $0.40$ GeV, respectively. The wave function is dominated by the ${}^2S_{1/2}$ component, with a small but non-negligible ${}^4D_{1/2}$ admixture that grows from $0.27\%$ to $0.74\%$ as the cutoff increases. When the coupled-channel calculation is performed including the $\Xi_b^* \bar B^*$ channel, the mixing is found to be extremely weak, and the $\Xi_b^* \bar B^*$ probability remains below $0.03\%$ across the entire cutoff range.

For the $\Xi_b^\prime \bar B^*$ state with $I(J^P)=0(3/2^-)$, the single-channel binding energies are $-2.32$, $-6.95$, and $-15.69$ MeV for $\Lambda = 0.25$, $0.33$, and $0.42$ GeV, respectively. The $D$-wave admixtures are slightly larger than in the $\Xi_b^\prime \bar B^*$ state with $I(J^P)=0(1/2^-)$, with the ${}^4D_{3/2}$ component reaching $1.38\%$ at the largest cutoff. Again, the coupled-channel mixing with the $\Xi_b^* \bar B^*$ channel is negligibly small, with the $\Xi_b^\prime \bar B^*$ component remaining above $99.9\%$ for all cutoff values.

For the $\Xi_b^* \bar B^*$ state with $I(J^P)=0(1/2^-)$, the binding energies are found to be $-2.13$, $-6.11$, and $-12.42$ MeV at the cutoff parameters $\Lambda = 0.27$, $0.34$, and $0.40$ GeV, respectively. The wave function is dominated by the ${}^2S_{1/2}$ partial wave, with tiny admixtures from the ${}^4D_{1/2}$ and ${}^6D_{1/2}$ components that gradually increase as the cutoff grows. These $D$-wave fractions remain below the percent level even at the largest cutoff, indicating that this loosely bound state preserves a nearly pure $S$-wave molecular nature.

For the $\Xi_b^* \bar B^*$ state with $I(J^P)=0(3/2^-)$, the binding energies are $-2.17$, $-6.75$, and $-13.87$ MeV at $\Lambda = 0.26$, $0.34$, and $0.41$ GeV, respectively. The dominant configuration is the ${}^4S_{3/2}$ component, while the $D$-wave admixtures stay below $1.3\%$ even at the highest cutoff, suggesting that the tensor-force effects are relatively mild in this bound state.

The most intriguing case is the $\Xi_b^* \bar B^*$ state with $I(J^P)=0(5/2^-)$, with the binding energies of $-2.07$, $-7.24$, and $-15.98$ MeV at $\Lambda = 0.24$, $0.33$, and $0.42$ GeV, respectively. As $\Lambda$ increases from $0.24$ GeV to $0.42$ GeV, the ${}^6D_{5/2}$ probability grows steadily from $0.34\%$ to $1.56\%$, reflecting the increasing role of the tensor interaction. Despite this growth, the $D$-wave admixture remains relatively small, so this loosely bound state still retains a predominantly $S$-wave molecular character while benefiting from the extra binding provided by the tensor force.

\begin{figure}[htbp]
\includegraphics[scale=0.60]{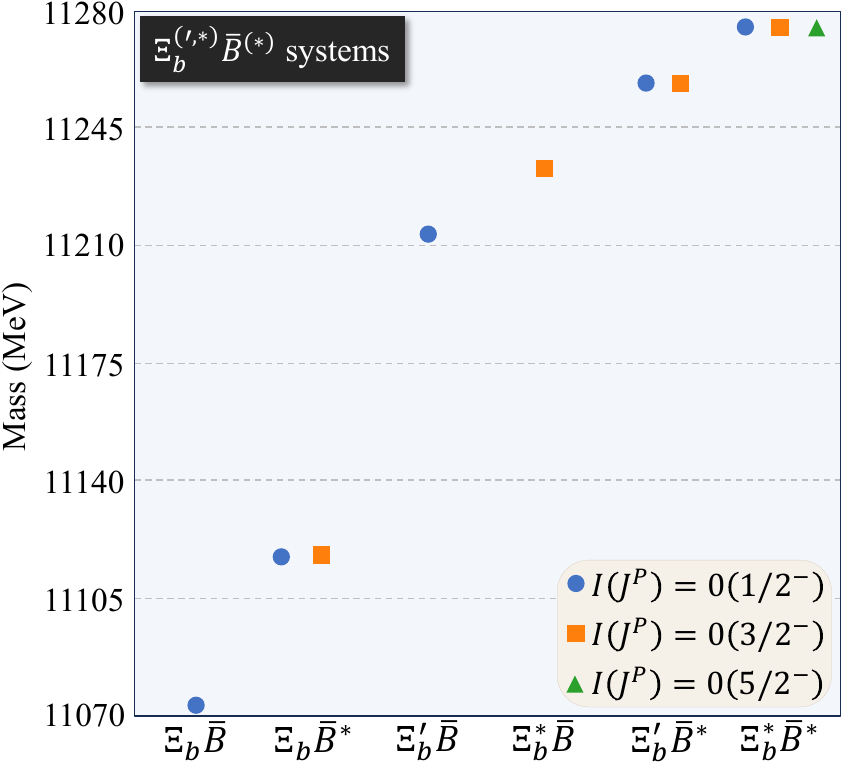}
\caption{Predicted mass spectra for the promising candidates of the $\Xi_b^{(\prime,\,*)} \bar B^{(*)}$-type singly-strange double-bottom molecular pentaquarks.}\label{XibbarBmassspectra}
\end{figure}

Based on our comprehensive analysis, we now summarize the promising candidates of the $\Xi_b^{(\prime,\,*)} \bar B^{(*)}$-type singly-strange double-bottom molecular pentaquarks, i.e.,
\begin{itemize}
\item $\Xi_b \bar B$ state with $I(J^P) = 0(1/2^-)$;
\item $\Xi_b \bar B^*$ states with $I(J^P) = 0(1/2^-)$ and $0(3/2^-)$;
\item $\Xi_b^\prime \bar B$ state with $I(J^P) = 0(1/2^-)$;
\item $\Xi_b^* \bar B$ state with $I(J^P) = 0(3/2^-)$;
\item $\Xi_b^\prime \bar B^*$ states with $I(J^P) = 0(1/2^-)$ and $0(3/2^-)$;
\item $\Xi_b^* \bar B^*$ states with $I(J^P) = 0(1/2^-)$, $0(3/2^-)$, and $0(5/2^-)$.
\end{itemize}
The corresponding mass spectra for these candidates are illustrated in Fig. \ref{XibbarBmassspectra}.

\section{Summary and Outlook}\label{summary}

In this work, we have carried out a systematic investigation of the mass spectra of
singly-strange molecular pentaquarks in both the hidden-bottom and double-bottom
sectors, focusing on the $\Xi_b^{(\prime,\,*)}B^{(*)}$ and
$\Xi_b^{(\prime,\,*)}\bar B^{(*)}$ systems. Our investigation is motivated by the
experimental observations of the singly-strange hidden-charm pentaquarks
$P_{\psi s}^{\Lambda}(4459)$ and $P_{\psi s}^{\Lambda}(4338)$, which suggest the
existence of bottom analogues through HQFS. Using the OBE model, we have derived the
effective potentials for the $\Xi_b^{(\prime,\,*)}B^{(*)}$ systems from effective
Lagrangians constrained by HQSS, chiral symmetry, and HLS. The effective potentials
for the $\Xi_b^{(\prime,\,*)}\bar B^{(*)}$ systems are then obtained through the
$G$-parity rule, which provides a consistent connection between the particle and
antiparticle channels. By solving the coupled-channel Schr\"odinger equation with
$S$--$D$ wave mixing and coupled-channel effects, we have searched for loosely bound
states with all possible isospin and spin-parity quantum numbers.

For the singly-strange hidden-bottom $\Xi_b^{(\prime,\,*)}B^{(*)}$ systems, we have
identified a set of promising molecular pentaquark candidates in the isoscalar
channels. These include the $\Xi_b B$ state with $I(J^P)=0(1/2^-)$, the $\Xi_b B^*$
states with $I(J^P)=0(1/2^-)$ and $0(3/2^-)$, the $\Xi_b^\prime B$ state with
$I(J^P)=0(1/2^-)$, the $\Xi_b^* B$ state with $I(J^P)=0(3/2^-)$, the
$\Xi_b^\prime B^*$ states with $I(J^P)=0(1/2^-)$ and $0(3/2^-)$, and the
$\Xi_b^* B^*$ states with $I(J^P)=0(1/2^-)$, $0(3/2^-)$, and $0(5/2^-)$. In contrast,
no loosely bound-state solutions are found in the isovector channels within the
reasonable cutoff range. The coupled-channel effects are found to be significant in
several systems. The most pronounced mixing occurs in the $\Xi_b^\prime B^*$ coupled
system with $I(J^P)=0(3/2^-)$, where the $\Xi_b^\prime B^*$ and $\Xi_b^* B^*$
components are nearly equally mixed. In contrast, the $\Xi_b B$ coupled system with
$I(J^P)=0(1/2^-)$ remains predominantly a single-channel state with minimal
admixtures, reflecting the suppressed off-diagonal transitions in this channel.

For the singly-strange double-bottom $\Xi_b^{(\prime,\,*)}\bar B^{(*)}$ systems, the
$G$-parity transformation modifies the effective interactions by flipping the signs
of the $\pi$ and $\omega$ exchange contributions while keeping those of the $\sigma$,
$\eta$, and $\rho$ unchanged. We have identified a set of promising singly-strange
double-bottom molecular pentaquark candidates in the isoscalar channels. These include
the $\Xi_b \bar B$ state with $I(J^P)=0(1/2^-)$, the $\Xi_b \bar B^*$ states with
$I(J^P)=0(1/2^-)$ and $0(3/2^-)$, the $\Xi_b^\prime \bar B$ state with
$I(J^P)=0(1/2^-)$, the $\Xi_b^* \bar B$ state with $I(J^P)=0(3/2^-)$, the
$\Xi_b^\prime \bar B^*$ states with $I(J^P)=0(1/2^-)$ and $0(3/2^-)$, and the
$\Xi_b^* \bar B^*$ states with $I(J^P)=0(1/2^-)$, $0(3/2^-)$, and $0(5/2^-)$. In
contrast, no loosely bound-state solutions are found in the isovector channels within
the reasonable cutoff range. The coupled-channel effects in the double-bottom sector
are generally less significant than those in the hidden-bottom sector. The most
pronounced difference occurs in the $\Xi_b^\prime \bar B^*$ coupled system with
$I(J^P)=0(3/2^-)$, where the mixing with the $\Xi_b^* \bar B^*$ channel is strongly
suppressed. This behavior is in sharp contrast to the hidden-bottom counterpart, in
which the two channels exhibit comparable mixing strengths. This suppression can be
traced to the significant reduction of the off-diagonal coupling matrix elements
induced by the $G$-parity transformation.

The predictions presented in this work provide concrete targets for future
experimental searches. At the LHCb experiment, the
$\Xi_b^{(\prime,\,*)}B^{(*)}$ and $\Xi_b^{(\prime,\,*)}\bar B^{(*)}$ molecular
pentaquark candidates can be produced in high-energy $pp$ collisions. With the
upgraded detector operating in Run III and the planned Run IV and High-Luminosity LHC
phases, LHCb is expected to accumulate datasets of $23\,\text{fb}^{-1}$,
$50\,\text{fb}^{-1}$, and ultimately $300\,\text{fb}^{-1}$, respectively
\cite{Bediaga:2018lhg,Dai:2026fkg}. This steady increase in integrated luminosity,
combined with improved trigger and reconstruction capabilities, will greatly enhance
the sensitivity to the predicted states. The experimental landscape for discovering
singly-strange hidden-bottom and double-bottom pentaquarks is therefore promising.

In conclusion, our analysis predicts a series of hidden-bottom and double-bottom
molecular pentaquark states with single strangeness, formed from the
$\Xi_b^{(\prime,\,*)}B^{(*)}$ and $\Xi_b^{(\prime,\,*)}\bar B^{(*)}$ interactions.
These states represent natural extensions of the observed singly-strange hidden-charm
molecular pentaquarks $P_{\psi s}^{\Lambda}(4459)$ and $P_{\psi s}^{\Lambda}(4338)$
into the bottom sector, and their experimental discovery would provide important
insights into hadron structure and the nonperturbative dynamics of strong
interactions.

\section*{Acknowledgement}

This work is supported by the Natural Science Foundation of Gansu Province (No. 26RCKA012 and No. 25JRRA799), the National Natural Science Foundation of China under Grants No. 12335001, No. 12405097, and No. 12247101, the ``111 Center" under Grant No. B20063, the fundamental Research Funds for the Central Universities (lzujbky-2023-stlt01), Lanzhou City High-Level Talent Funding, and the Talent Scientific Fund of Lanzhou University. R. Chen is also supported by the National Natural Science Foundation of China under Grant No. 12305139, and the Xiaoxiang Scholars Programme of Hunan Normal University. 

\end{document}